\documentclass[aps, prl, reprint, superscriptaddress]{revtex4-2}
\usepackage{hyperref}
\usepackage{xcolor}
\usepackage{physics}
\usepackage{amsfonts}
\usepackage{bbm}
\usepackage{amsmath}
\usepackage{graphicx}

\newcommand{\im}{\mathrm{i}}

\begin{document}


\title{Detecting non-Hermitian topology from long-time dynamics}

\date{\today}

\author{Antonio D'Abbruzzo}
\affiliation{The Abdus Salam International Centre for Theoretical Physics, Strada Costiera 11, I-34151 Trieste, Italy}

\author{Rosario Fazio}
\affiliation{The Abdus Salam International Centre for Theoretical Physics, Strada Costiera 11, I-34151 Trieste, Italy}
\affiliation{Dipartimento di Fisica, Università di Napoli ``Federico II'', Monte S. Angelo, I-80126, Napoli, Italy}

\begin{abstract}
We propose a protocol to measure the non-Hermitian topological invariant associated with the Lindbladian spectrum of chiral-symmetric open Gaussian systems from the relaxation of the covariance matrix.
Importantly, the method works in the long-time regime and does not rely on isolating edge modes, so it remains applicable even when the Liouvillian skin effect may obscure them.
We illustrate the technique on a Kitaev chain with reservoir-induced nonreciprocity, for which we first obtain a fully analytical and previously unknown description of its topological phase diagram.
We then show that our protocol reliably distinguishes all the different phases.
\end{abstract}

\maketitle


\textit{Introduction---}%
The study of topological effects in closed quantum systems constitutes a well-established research field in condensed matter physics~\cite{altland1997symmetry, bernevig2013book, chiu2016classification}.
Due to potential applications in the context of quantum technologies, considerable attention has been given in recent years to extending topological theories to open quantum systems.
For what concerns Lindblad dynamics~\cite{lindblad1976original, gorini1976original, breuer2007book, rivas2012book}, one mainly finds two independent approaches: mixed-state topology of the density matrix~\cite{bardyn2013topology, viyuela2014uhlmann, bardyn2018probing, altland2021symmetry} and non-Hermitian topology of the Lindbladian spectrum~\cite{vancaspel2019dynamical, lieu2020tenfold, sa2023symmetry, mao2025symmetry, chaduteau2026postselect}.
Symmetry classifications have been obtained in both cases~\cite{altland2021symmetry, lieu2020tenfold, sa2023symmetry, mao2025symmetry}, finding several interesting effects, such as dissipative edge states~\cite{zhu2021delocalization, wang2022morphing}, Liouvillian skin effect~\cite{yang2022skin, zhou2022adjoint, lin2023skin}, relaxation slowdown without gap closing~\cite{haga2021slowing}, and chiral damping~\cite{song2019chiral}.

While the exploration of mixed-state topology is performed directly in terms of the state, linking the spectrum topology to measurable properties of the density matrix is less straightforward because it is encoded into the relaxation towards the steady state rather than the steady state itself.
Techniques based on the detection of topological edge modes have been recently proposed~\cite{sarkar2026chiral, yang2026boundary}, even though their effectiveness is limited in the presence of skin effect, where such modes may spatially or spectrally conflate with the others.
A more direct topology probe was introduced in the form of entanglement spectrum crossings~\cite{sayyad2021entanglement, bayona2025entanglement}, but it relies on detecting events happening at finite times and, additionally, it requires a dissipation structure that may be experimentally challenging to implement.
Therefore, a reliable and general framework to detect non-Hermitian topology in open quantum systems is arguably still missing.

In this work, focusing on one-dimensional chiral-symmetric models, we devise a protocol to measure the topological invariant directly from the long-time behavior of the covariance matrix, thereby bridging the gap between Lindbladian spectrum topology and density matrix.
As we will show, the method is based on the fact that the ratio of carefully chosen components of the covariance matrix converges at long time to a topology-aware quantity, regardless of the steady state reached by the components themselves.
This is similar in spirit to the works employing spin textures for detecting the topology of non-Hermitian Hamiltonians~\cite{wang2021detecting, lin2025quench}.

As an example, we further analyze a Kitaev chain~\cite{kitaev2001original} featuring reservoir-induced nonreciprocal dynamics~\cite{metelmann2015nonreciprocal, mcdonald2022nonreciprocal, soares2026nonreciprocal}, which exhibits both dissipative topological edge modes and Liouvillian skin effect.
Even though such model was already considered before~\cite{brighi2026kitaev}, here we obtain for the first time a fully analytical and previously unknown construction of its topological phase diagram, revealing how the transition point is altered by the dissipation.
Then, we provide a numerical demonstration that our protocol is able to reliably distinguish all the different phases.


\textit{Setup---}%
Let us consider a one-dimensional lattice of $N$ noninteracting fermions, linearly coupled to a Markovian environment.
The evolution of the density matrix $\hat\rho$ is dictated by a Lindblad equation~\cite{lindblad1976original, gorini1976original, breuer2007book, rivas2012book}:
\begin{equation} \label{eq:lindblad}
	\dv{\hat\rho}{t} = -\im[\hat H, \hat\rho] + \sum_{\alpha=1}^M \qty[
	    \hat L_\alpha \hat\rho \hat L_\alpha^\dagger
		- \frac{1}{2} \qty{\hat L_\alpha^\dagger \hat L_\alpha, \hat\rho}
	],
\end{equation}
where $\hat H$ is a Hamiltonian operator and $\hat{\mathbf L} = (\hat L_\alpha)_{\alpha=1}^M$ is a vector of $M$ Lindblad operators.
We set units in which $\hbar = 1$ and we reserve the hat notation for operators acting on the system's Hilbert space.
For each lattice site index $j \in \qty{1, \ldots, N}$, let $\hat a_j$, $\hat a_j^\dagger$ be the ladder operators associated with it.
By hypothesis, $\hat H$ is quadratic and each $\hat L_\alpha$ is linear in such operators~\cite{dabbruzzo2021self}.
Following Refs.~\cite{prosen2008third, barthel2022solving}, it is more convenient to express them in terms of the vector $\hat{\mathbf w}^\dagger = (\hat w_1^A, \dots, \hat w_N^A, \hat w_1^B, \dots, \hat w_N^B)$ of $2N$ Majorana operators $\hat w_j^A := (\hat a_j^\dagger + \hat a_j)/\sqrt{2}$ and $\hat w_j^B := \im (\hat a_j^\dagger - \hat a_j)/\sqrt{2}$.
We then assume $\hat H = \frac{\im}{2} \hat{\mathbf w}^\dagger H \hat{\mathbf w}$, with $H$ being a $2N \times 2N$ real skew-symmetric matrix, and $\hat{\mathbf L} = L \hat{\mathbf w}$, with $L$ being a $M \times 2N$ complex matrix.
Because of this structure, $\hat\rho$ is completely determined by the covariance matrix $\Gamma := -\im \Tr([\hat{\mathbf w}, \hat{\mathbf w}^\dagger] \hat\rho)$, which obeys~\cite{barthel2022solving}
\begin{equation} \label{eq:lyapunov}
	\dv{\Gamma}{t} = X \Gamma + \Gamma X^T + Y,
\end{equation}
where $X := H - \Re(L^\dagger L)$ is the damping matrix, and $Y := 2 \Im(L^\dagger L)$.
We assume the steady state $\Gamma_\infty$ reached for $t \to \infty$ to be unique~\cite{nigro2019uniqueness}.
Note that the damping matrix does not fully determine $\Gamma_\infty$, but it does determine the relaxation dynamics towards it.
In fact, if we define the covariance deviation $G := \Gamma - \Gamma_\infty$, we find
\begin{equation} \label{eq:lyapunov_g}
	\dv{G}{t} = X G + G X^T
	\quad \Rightarrow \quad
	G(t) = e^{X t} G(0) e^{X^T t}.
\end{equation}

Since $X$ is not symmetric, it generically has complex spectrum and non-Hermitian topology theory applies to it.
Several works addressed the classification and the definition of topological invariants for generic non-Hermitian matrices~\cite{bernard2002classification, kawabata2019symmetry, lieu2020tenfold, kawasaki2022shifted, mao2025symmetry}.
In this work, we focus on the very relevant class obeying the symmetry
\begin{equation} \label{eq:chiral_symmetry}
	X + \gamma \mathbbm{1}_{2N} = - U \qty[X + \gamma \mathbbm{1}_{2N}] U^{-1},
\end{equation}
where $U$ is a unitary matrix satisfying $U^2 = \mathbbm{1}_{2N}$ and $\gamma$ is fixed by the condition $\Tr[X + \gamma \mathbbm{1}_{2N}] = 0$.
Recall that $\Tr X \leq 0$, so that $\gamma \geq 0$.
Eq.~\eqref{eq:chiral_symmetry} was called shifted sublattice symmetry in Ref.~\cite{kawasaki2022shifted}, and is the natural open-system analogue of the chiral symmetry that is usually defined for closed systems.
In case the bulk of $X$ is translationally invariant, the $2 \times 2$ blocks pertaining to spatial indices $i$ and $j$ actually depend only on the difference $i-j$: after writing $n = i-j$, we will use the notation $X^{(n)}$ to indicate such blocks.
The topology of $X$ is then encoded into the so-called symbol $\mathcal{X}(z) := \sum_{n \in \mathbb{Z}} X^{(n)} z^n$, with $z \in \mathbb{C} \setminus \qty{0}$~\cite{bottcher1999book}.
Calligraphic font will be used in the following to refer to symbols.
Similarly to the closed-system scenario, one can prove that the symbol of a matrix obeying the symmetry~\eqref{eq:chiral_symmetry} is constrained to have the form~\cite{kawabata2019symmetry, yokomizo2019gbz, kawasaki2022shifted}
\begin{equation} \label{eq:symbol}
	\mathcal{X}(z) = \begin{pmatrix}
		-\gamma & \alpha(z) \\
		\beta(z) & -\gamma
	\end{pmatrix},
\end{equation}
with $\alpha(z)$ and $\beta(z)$ being complex Laurent polynomials.
Moreover, its point-gap topological phase under open boundary conditions is characterized by the quantized winding number~\cite{kawabata2019symmetry, yokomizo2019gbz}
\begin{equation} \label{eq:winding}
	\nu := \frac{1}{4 \pi \im} \oint_\text{GBZ} \dd{\log \qty(\frac{\alpha(z)}{\beta(z)})}.
\end{equation}
The integral is performed along the generalized Brillouin zone (GBZ), a deformation of the standard Brillouin zone into complex momentum space designed to restore the bulk-boundary correspondence in the presence of skin effect~\cite{yokomizo2019gbz, yang2020auxiliary, wang2026gbz}.
In order for $\nu$ to be well defined, we require $\alpha(z)$ and $\beta(z)$ to be nonzero for $z$ ranging throughout the contour of interest.


\textit{The protocol---}%
We argue that it is possible to determine the winding number $\nu$ by only having access to the long-time behavior of the covariance deviation $G(t)$.

In order to explain the basic idea more easily, we start with the assumption of infinite system $N \to \infty$, so that we can ignore boundary effects.
We will discuss later how to properly treat finite-size systems.
Moreover, assume $X$ and $Y$ to be translationally invariant.
One can verify~\cite{sm} that the solution $\Gamma(t)$ to Eq.~\eqref{eq:lyapunov} remains translationally invariant at all times $t \geq 0$, provided one starts with a translationally invariant initial condition $\Gamma(0)$.
Consequently, the deviation $G(t)$ is translationally invariant too throughout the whole evolution.

From Eq.~\eqref{eq:lyapunov_g} we can see that the following evolution holds for the blocks of $G(t)$:
\begin{equation}
    \dv{G^{(n)}}{t} = \sum_{m \in \mathbb{Z}} \qty[
        X^{(m)} G^{(n-m)} + G^{(m)} (X^{(m-n)})^T
    ].
\end{equation}
Defining the symbol $\mathcal{G}(z, t) := \sum_{n \in \mathbb{Z}} G^{(n)}(t) z^n$, we have
\begin{equation} \label{eq:sylvester_G}
	\dv{\mathcal{G}(z, t)}{t} = \mathcal{X}(z) \mathcal{G}(z, t) + \mathcal{G}(z, t) \mathcal{X}^T (z^{-1}),
\end{equation}
with solution
\begin{equation} \label{eq:G}
	\mathcal{G}(z, t) = e^{\mathcal{X}(z) t} \mathcal{G}(z, 0) e^{\mathcal{X}^T(z^{-1}) t}.
\end{equation}
In order to ease the notation, let us temporarily drop the $z$ dependence by writing, e.g., $\mathcal{G}(t) \equiv \mathcal{G}(z, t)$ and $\mathcal{X} \equiv \mathcal{X}(z)$.
We also write $\mathcal{X}' \equiv \mathcal{X}^T(z^{-1})$.
Any quantity defined for $\mathcal{X}$ will also be immediately defined for $\mathcal{X}'$ by adding primes.
The matrix $\mathcal{X}$ has the spectral decomposition
\begin{equation} \label{eq:symbol_spectral}
	\mathcal{X} + \gamma \mathbbm{1}_2 = P \begin{pmatrix}
		\varphi & 0 \\
		0 & -\varphi
	\end{pmatrix} P^{-1},
	\qquad
	P = \begin{pmatrix}
		\alpha & \alpha \\
		\varphi & -\varphi
	\end{pmatrix},
\end{equation}
where $\varphi := \sqrt{\alpha\beta}$.
Using Eq.~\eqref{eq:symbol_spectral} into Eq.~\eqref{eq:G}, we find the following expression for the matrix elements of $\mathcal{G}(t)$:
\begin{equation} \label{eq:G_decomposed}
	\mathcal{G}_{ab}(t) = e^{-2 \gamma t} \sum_{c,d=1}^2 e^{(\varphi_c + \varphi'_d) t}
	P_{ac} F_{cd} P'^{-1}_{db},
\end{equation}
where $F := P^{-1} \mathcal{G}(0) P'$ and $\varphi_{1,2} := \pm \varphi$.
Let us now assume we are away from the branch cut of the principal square root used to define $\varphi$, meaning that $\alpha\beta \notin (-\infty, 0]$ and $\Re \varphi > 0$.
Similarly, we assume $\Re \varphi' > 0$.
Then, we can identify a long-time dominant contribution:
\begin{equation} \label{eq:G_decomposed2}
	\mathcal{G}_{ab}(t) = e^{-2 \gamma t} e^{(\varphi + \varphi')t} \qty[
	    P_{a 1} F_{11} P'^{-1}_{1 b} + \order{e^{-\delta t}}
	]
\end{equation}
with $\delta := 2 \min \qty{\Re \varphi, \Re \varphi'} > 0$.
Provided the initial condition satisfies $F_{11} \neq 0$, the ratio between $\mathcal{G}_{11}(t)$ and $\mathcal{G}_{21}(t)$ therefore converges exponentially fast to a finite limit at long time:
\begin{equation} \label{eq:ratio}
	\qty[ \frac{\mathcal{G}_{11}(t)}{\mathcal{G}_{21}(t)} ]^2
	\xrightarrow{t \to \infty}
	\qty[ \frac{P_{11}}{P_{21}} ]^2 = \frac{\alpha}{\beta}.
\end{equation}
Recall that $G(t) \to 0$ for $t \to \infty$, so that $\mathcal{G}_{11}(t)$ and $\mathcal{G}_{21}(t)$ individually decay to zero.
Interestingly, their ratio instead becomes aligned to $\alpha/\beta$, which is precisely the function of which we have to perform the winding to get the topological invariant $\nu$ in Eq.~\eqref{eq:winding}.

An important aspect to understand is how the previous analysis is affected by finite-size effects.
Non-Hermitian matrices are notoriously sensitive to boundary conditions~\cite{trefethen2005book} and they matter here too.
For a finite-size system with periodic boundary conditions, it can be shown that Eqs.~\eqref{eq:sylvester_G}--\eqref{eq:ratio} continue to be valid provided one restricts $z$ to the Fourier points $z_k := e^{2 \pi \im k/N}$, $k=1,\ldots,N$, so that Eq.~\eqref{eq:ratio} can still be used to probe $\alpha/\beta$ on such points~\cite{sm}.
Assuming $N$ to be sufficiently large, it is then possible to numerically reconstruct $\alpha/\beta$ for arbitrary $z$ in an efficient manner using rational approximation algorithms~\cite{nakatsukasa2018aaa, sm}.

According to Eq.~\eqref{eq:winding}, it seems now that one needs to sweep $z$ along the GBZ in order to recover the correct invariant.
This is not desirable because the GBZ highly depends on the details of the functions $\alpha(z)$ and $\beta(z)$~\cite{yokomizo2019gbz}, a fact that would hinder a model-agnostic application of the method.
We now show how to bypass this issue.

First, it is actually possible to infer $\nu$ exclusively from the polynomial structure of $\alpha(z)$ and $\beta(z)$.
Let $R > 0$ be the highest negative power of $z$ appearing in $\alpha(z)$ and $\beta(z)$, which is basically the interaction range of the damping matrix~\cite{yokomizo2019gbz}.
Also, let $\qty{z_i}$ be the set of roots and poles of $\alpha(z)/\beta(z)$ ordered by magnitude, $|z_i| \leq |z_{i+1}|$.
Then, Refs.~\cite{lee2019anatomy, zhong2024pole} established that the GBZ always encloses only the first $2R$ of them, $z_1, \ldots, z_{2R}$.
By Cauchy's argument principle~\cite{sveshnikov1978book} this is enough to infer $\nu$: it is the half difference between the number of roots and poles in the set $\qty{z_1, \ldots, z_{2R}}$.
We can then apply our method by only knowing $R$ and not the specific values of $z_1, \ldots, z_{2R}$.
In fact, consider the winding $\nu(r)$ of $\alpha(z)/\beta(z)$ along a circle of radius $r$ [Eq.~\eqref{eq:winding} with GBZ replaced by $|z| = r$].
This quantity can be estimated using the protocol above without knowing the details of the model.
If we sweep the value of $r$, we observe jumps in $\nu(r)$ precisely when the circle intersects one of the $z_i$.
Provided we start the sweep from a sufficiently small $r$, where $\nu(r) = 0$, we can therefore obtain the topological invariant $\nu$ in~\eqref{eq:winding} by reading the value of $\nu(r)$ after $2R$ jumps.

The protocol fails to probe a certain $z$ in case $\Re \varphi = 0$ or $\Re \varphi' = 0$.
In this case, it is not possible to identify a long-time dominant contribution to Eq.~\eqref{eq:G_decomposed} and the ratio oscillates indefinitely.
However, since $z$ is ultimately swept on circles, the product $\alpha(z) \beta(z)$ typically yields loops in the complex plane and the pathological points are few and isolated.
Practically, they can be avoided without spoiling the calculation of $\nu(r)$, which does not require accurate probing over the whole circle~\cite{sm}.

The protocol we propose can finally be summarized as follows.
One starts from the (possibly experimentally determined) two-point correlation functions $\mathrm{Tr}[\hat a_i^\dagger \hat a_j \hat\rho(t)]$ and $\mathrm{Tr}[\hat a_i \hat a_j \hat \rho(t)]$, for each $i, j \in \qty{1, \ldots, N}$, and constructs the Majorana covariance $\Gamma(t)$.
Given the steady state $\Gamma_\infty$, one considers the deviation $G(t) = \Gamma(t) - \Gamma_\infty$ and calculates its discrete Fourier transform $\mathcal{G}(z_k, t)$.
For each $k \in \qty{1, \ldots, N}$ one extracts the long-time limit $f(z_k) := \lim_{t \to \infty} [\mathcal{G}_{11}(z_k, t) / \mathcal{G}_{21}(z_k, t)]^2$ and then extrapolates $f(z)$ to arbitrary $z$, in order to be able to calculate its winding $\nu(r)$ on circles of radius $r$.
Finally, sweeping $r$ and reading $\nu(r)$ after $2R$ jumps yields the desired topological invariant $\nu$.


\begin{figure*}
	\includegraphics[width=\linewidth]{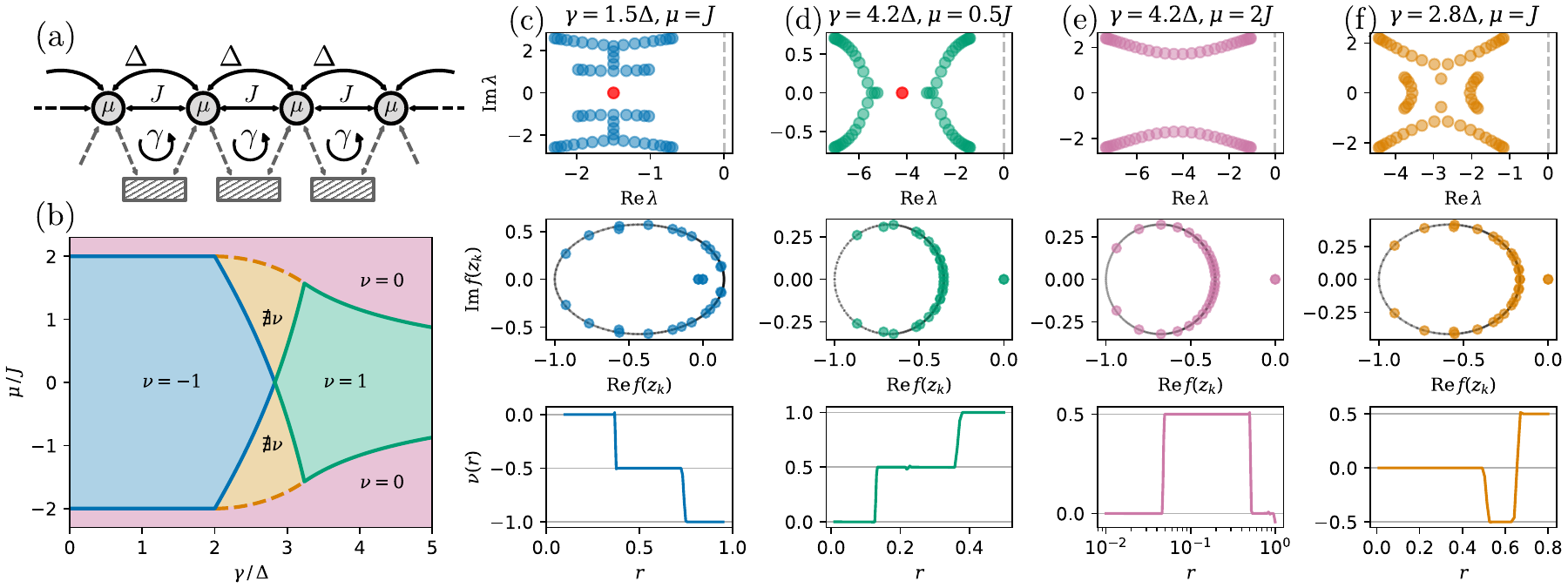}
	\caption{\label{fig}
	    (a) Schematic depiction of the nonreciprocal Kitaev chain defined by Eqs.~\eqref{eq:kitaev_hamiltonian}--\eqref{eq:kitaev_lindblads}.
		(b) Topological phase diagram of the model, plotted for $J = \Delta$; its exact analytical description is given in Eqs.~\eqref{eq:critical}--\eqref{eq:exceptional}.
		(c-f) Numerical demonstration of the protocol for four choices of parameters lying in the four regions identified in (b).
		Top row: spectrum of $X$ under open boundary conditions with $N=30$; topological edge modes, when present, are highlighted in red.
		Middle row: example of ratios $f(z_k) := [\mathcal{G}_{11}(z_k, t)/\mathcal{G}_{21}(z_k, t)]^2$ at the Fourier points $z_k := e^{2 \pi \im k/N}$, $k=1,\ldots,N$, calculated under periodic boundary conditions for $N = 30$ and $\gamma t = 10$, with the initial state being the fully occupied configuration; the thin gray points indicate the analytical prediction $\alpha(z_k)/\beta(z_k)$.
		Bottom row: winding number $\nu(r)$ of $f(z)$ as a function of the radius $r$; the topological invariant $\nu$ of Eq.~\eqref{eq:winding} equals $\nu(r)$ after two jumps occurred ($R = 1$).
		In the exceptional region (f) $\nu$ is not defined and the protocol yields the value $1/2$.
	}
\end{figure*}

\textit{Nonreciprocal Kitaev chain---}%
As a paradigm model, we consider the Kitaev chain~\cite{kitaev2001original} schematically depicted in Fig.~\ref{fig}(a).
The Hamiltonian is
\begin{equation} \label{eq:kitaev_hamiltonian}
	\hat H = \mu \sum_j \hat a_j^\dagger \hat a_j
	+ \sum_j \qty[J \hat a_j^\dagger \hat a_{j+1} + \Delta \hat a_j^\dagger \hat a_{j+1}^\dagger + \text{H.c.}],
\end{equation}
and we take a Lindblad operator for each site, $\hat{L}_\alpha \equiv \hat{L}_j$, according to
\begin{equation} \label{eq:kitaev_lindblads}
	\hat L_j = \sqrt{\gamma} \qty(\hat a_j + \im \, \hat a_{j+1}).
\end{equation}
This choice of dissipation constitutes the standard pathway to effective nonreciprocal hoppings in the covariance dynamics~\cite{metelmann2015nonreciprocal, mcdonald2022nonreciprocal, soares2026nonreciprocal}, which is different from studying non-Hermitian extensions of the Kitaev chain~\cite{chaduteau2026postselect, zhao2021defective, ardonne2025kitaev, ray2026kitaev}.
For simplicity, we here assume the parameters to be real, and $J, \Delta, \gamma > 0$.
Once we move to Majorana operators, $\hat H$ and $\hat L_j$ have, respectively, the quadratic and linear shape assumed below Eq.~\eqref{eq:lindblad}.
A straightforward computation shows that the damping matrix $X$ of this model satisfies the symmetry~\eqref{eq:chiral_symmetry}, and that the symbol $\mathcal{X}(z)$ is given by Eq.~\eqref{eq:symbol} with
\begin{subequations}
\begin{align}
	\alpha(z) &= (J_- + \Delta) z^{-1} + \mu + (J_+ - \Delta) z, \\
	\beta(z) &= -(J_- - \Delta) z^{-1} - \mu - (J_+ + \Delta) z,
\end{align}
\end{subequations}
where $J_\pm := J \pm \gamma/2$~\cite{sm}.
The constant $\gamma$---representing both dissipation rate and degree of nonreciprocity---is the same that appears on the diagonal of $\mathcal{X}(z)$ in Eq.~\eqref{eq:symbol}.
Also, the interaction range is $R = 1$.

The point-gap topological phases of this model can be studied exactly: as we said previously, $\nu$ is determined by the polynomial structure of $\alpha(z)/\beta(z)$, whose roots and poles can be here written explicitly.
The full computation is however quite involved: we report the final result in the End Matter [Eqs.~\eqref{eq:critical}--\eqref{eq:exceptional}] and its derivation in the Supplemental Material~\cite{sm}.
Here we depict the obtained phase diagram in Fig.~\ref{fig}(b).
We find three regions characterized by $\nu \in \qty{-1, 0, 1}$.
When $|\nu| = 1$ the corresponding open chain exhibits a pair of edge modes, with $\mathrm{sgn}(\nu)$ determining the Majorana chirality of the localization.
We also find an extended ``exceptional region'' where the eigenvectors of $\mathcal{X}(z)$, for a certain $z$ on the GBZ, coalesce and $\nu$ ceases to be well defined~\cite{yokomizo2019gbz}.
Furthermore, note that the standard Kitaev transition at $|\mu| = 2J$~\cite{kitaev2001original, dabbruzzo2021kitaev} remains unaffected by the dissipation for $\gamma \leq 2 \Delta$.
Hence, if $\gamma$ is sufficiently small, the protocol described here provides an alternative way to probe the ground-state topology of the Kitaev chain.

In Fig.~\ref{fig}(c-f) we showcase numerical results obtained by applying the protocol to a finite-size chain under periodic boundary conditions, with the initial state being the fully occupied configuration.
At the Fourier points $z_k$, we observe how the ratio $f(z_k, t) := [\mathcal{G}_{11}(z_k, t)/\mathcal{G}_{21}(z_k, t)]^2$ approaches at long (but finite) time the analytical prediction $\alpha(z_k)/\beta(z_k)$ except for few of such points, where either no dominant contribution to the ratio emerges or the initial state is fine-tuned to kill it.
Nevertheless, the whole $f(z)$ can be reconstructed using rational approximation and the windings $\nu(r)$ can be calculated from the resulting data.
We leave additional technical details about how the computation is practically carried out to the Supplemental Material~\cite{sm}.
The final result visible from Fig.~\ref{fig} is that choosing parameters in the three regions having $\nu \in \qty{-1, 0, 1}$, the method correctly retrieves the topological invariant.
In the exceptional region, where such invariant is not mathematically defined, the procedure yields instead a recognizable fractional value of $1/2$.


\textit{Conclusions---}%
In summary, we proposed a technique to measure the point-gap topological invariant of one-dimensional chiral-symmetric open Gaussian systems from the long-time relaxation of the covariance matrix.
We believe it provides an important step forward in understanding how the spectrum topology of an open system influences measurable properties of the state.
A natural future direction would be to understand whether such idea can be applied to multiband or multidimensional systems, or extended to other symmetry classes.


\textit{Acknowledgments---}%
This work was supported by the European Union (ERC RAVE, Grant ID: 101053159).
We thank Lorenzo Cirigliano, Giuseppe Santoro, and Aashish Clerk for useful discussions.
Claude Sonnet 5.5 LLM was used to brainstorm approaches to certain calculations contained in the Supplemental Material~\cite{sm}.


\section{End Matter}

The topological phase diagram of the nonreciprocal Kitaev chain defined by Eqs.~\eqref{eq:kitaev_hamiltonian}--\eqref{eq:kitaev_lindblads} is shown in Fig.~\ref{fig}(a).
Here we report the exact formulas that define that diagram.
The critical line $\mu_c$ separating the regions with $\nu = 0$ ($|\mu| > |\mu_c|$) and $|\nu| = 1$ ($|\mu| < |\mu_c|$) is
\begin{equation} \label{eq:critical}
	|\mu_c| = \begin{cases}
		2J & 0 \leq \gamma \leq 2\Delta, \\
		\frac{|\gamma^2 - 4(J^2 + \Delta^2)|}{\sqrt{4 J^2 - (\gamma - 2\Delta)^2}} & 2 \Delta \leq \gamma \leq \Delta + \sqrt{\Delta^2 + 4 J^2}, \\
		\frac{4 J \Delta}{\sqrt{\gamma^2 - 4 J^2}} & \gamma \geq \Delta + \sqrt{\Delta^2 + 4 J^2}.
	\end{cases}
\end{equation}
The point where the regions $\nu = 1$ and $\nu = -1$ intersect corresponds to $\gamma_c = 2 \sqrt{J^2 + \Delta^2}$.
Specifically, in the topologically nontrivial sector, $\nu = -1$ holds for $\gamma < \gamma_c$ and $\nu = 1$ holds for $\gamma > \gamma_c$.
The exceptional region, where $\nu$ does not exist, corresponds instead to
\begin{equation} \label{eq:exceptional}
	\frac{|\gamma^2 - 4(J^2 + \Delta^2)|}{\sqrt{4 J^2 - (\gamma - 2\Delta)^2}}
	\leq |\mu| \leq \sqrt{4 J^2 - (\gamma - 2\Delta)^2}
\end{equation}
for $2\Delta \leq \gamma \leq \Delta + \sqrt{\Delta^2 + 4 J^2}$.
In such range for $\gamma$, Eq.~\eqref{eq:exceptional} describes the intersection of $|\mu| > |\mu_c|$ with the circle of radius $2J$ centered in $\mu = 0$ and $\gamma = 2\Delta$.

The proof of all these statements can be found in the Supplemental Material~\cite{sm}.


\bibliography{biblio.bib}

\begin{thebibliography}{55}%
\makeatletter
\providecommand \@ifxundefined [1]{%
 \@ifx{#1\undefined}
}%
\providecommand \@ifnum [1]{%
 \ifnum #1\expandafter \@firstoftwo
 \else \expandafter \@secondoftwo
 \fi
}%
\providecommand \@ifx [1]{%
 \ifx #1\expandafter \@firstoftwo
 \else \expandafter \@secondoftwo
 \fi
}%
\providecommand \natexlab [1]{#1}%
\providecommand \enquote  [1]{``#1''}%
\providecommand \bibnamefont  [1]{#1}%
\providecommand \bibfnamefont [1]{#1}%
\providecommand \citenamefont [1]{#1}%
\providecommand \href@noop [0]{\@secondoftwo}%
\providecommand \href [0]{\begingroup \@sanitize@url \@href}%
\providecommand \@href[1]{\@@startlink{#1}\@@href}%
\providecommand \@@href[1]{\endgroup#1\@@endlink}%
\providecommand \@sanitize@url [0]{\catcode `\\12\catcode `\$12\catcode
  `\&12\catcode `\#12\catcode `\^12\catcode `\_12\catcode `\%12\relax}%
\providecommand \@@startlink[1]{}%
\providecommand \@@endlink[0]{}%
\providecommand \url  [0]{\begingroup\@sanitize@url \@url }%
\providecommand \@url [1]{\endgroup\@href {#1}{\urlprefix }}%
\providecommand \urlprefix  [0]{URL }%
\providecommand \Eprint [0]{\href }%
\providecommand \doibase [0]{https://doi.org/}%
\providecommand \selectlanguage [0]{\@gobble}%
\providecommand \bibinfo  [0]{\@secondoftwo}%
\providecommand \bibfield  [0]{\@secondoftwo}%
\providecommand \translation [1]{[#1]}%
\providecommand \BibitemOpen [0]{}%
\providecommand \bibitemStop [0]{}%
\providecommand \bibitemNoStop [0]{.\EOS\space}%
\providecommand \EOS [0]{\spacefactor3000\relax}%
\providecommand \BibitemShut  [1]{\csname bibitem#1\endcsname}%
\let\auto@bib@innerbib\@empty
\bibitem [{\citenamefont {Altland}\ and\ \citenamefont
  {Zirnbauer}(1997)}]{altland1997symmetry}%
  \BibitemOpen
  \bibfield  {author} {\bibinfo {author} {\bibfnamefont {A.}~\bibnamefont
  {Altland}}\ and\ \bibinfo {author} {\bibfnamefont {M.~R.}\ \bibnamefont
  {Zirnbauer}},\ }\bibfield  {title} {\bibinfo {title} {Nonstandard symmetry
  classes in mesoscopic normal-superconducting hybrid structures},\ }\href
  {https://doi.org/10.1103/PhysRevB.55.1142} {\bibfield  {journal} {\bibinfo
  {journal} {Phys. Rev. B}\ }\textbf {\bibinfo {volume} {55}},\ \bibinfo
  {pages} {1142} (\bibinfo {year} {1997})}\BibitemShut {NoStop}%
\bibitem [{\citenamefont {Bernevig}\ and\ \citenamefont
  {Hughes}(2013)}]{bernevig2013book}%
  \BibitemOpen
  \bibfield  {author} {\bibinfo {author} {\bibfnamefont {B.~A.}\ \bibnamefont
  {Bernevig}}\ and\ \bibinfo {author} {\bibfnamefont {T.~L.}\ \bibnamefont
  {Hughes}},\ }\href {https://doi.org/10.1515/9781400846733} {\emph {\bibinfo
  {title} {Topological Insulators and Topological Superconductors}}}\ (\bibinfo
   {publisher} {Princeton University Press},\ \bibinfo {year}
  {2013})\BibitemShut {NoStop}%
\bibitem [{\citenamefont {Chiu}\ \emph {et~al.}(2016)\citenamefont {Chiu},
  \citenamefont {Teo}, \citenamefont {Schnyder},\ and\ \citenamefont
  {Ryu}}]{chiu2016classification}%
  \BibitemOpen
  \bibfield  {author} {\bibinfo {author} {\bibfnamefont {C.-K.}\ \bibnamefont
  {Chiu}}, \bibinfo {author} {\bibfnamefont {J.~C.~Y.}\ \bibnamefont {Teo}},
  \bibinfo {author} {\bibfnamefont {A.~P.}\ \bibnamefont {Schnyder}},\ and\
  \bibinfo {author} {\bibfnamefont {S.}~\bibnamefont {Ryu}},\ }\bibfield
  {title} {\bibinfo {title} {Classification of topological quantum matter with
  symmetries},\ }\href {https://doi.org/10.1103/RevModPhys.88.035005}
  {\bibfield  {journal} {\bibinfo  {journal} {Rev. Mod. Phys.}\ }\textbf
  {\bibinfo {volume} {88}},\ \bibinfo {pages} {035005} (\bibinfo {year}
  {2016})}\BibitemShut {NoStop}%
\bibitem [{\citenamefont {Lindblad}(1976)}]{lindblad1976original}%
  \BibitemOpen
  \bibfield  {author} {\bibinfo {author} {\bibfnamefont {G.}~\bibnamefont
  {Lindblad}},\ }\bibfield  {title} {\bibinfo {title} {On the generators of
  quantum dynamical semigroups},\ }\href {https://doi.org/10.1007/BF01608499}
  {\bibfield  {journal} {\bibinfo  {journal} {Comm. Math. Phys.}\ }\textbf
  {\bibinfo {volume} {48}},\ \bibinfo {pages} {119} (\bibinfo {year}
  {1976})}\BibitemShut {NoStop}%
\bibitem [{\citenamefont {Gorini}\ \emph {et~al.}(1976)\citenamefont {Gorini},
  \citenamefont {Kossakowski},\ and\ \citenamefont
  {Sudarshan}}]{gorini1976original}%
  \BibitemOpen
  \bibfield  {author} {\bibinfo {author} {\bibfnamefont {V.}~\bibnamefont
  {Gorini}}, \bibinfo {author} {\bibfnamefont {A.}~\bibnamefont
  {Kossakowski}},\ and\ \bibinfo {author} {\bibfnamefont {E.~C.~G.}\
  \bibnamefont {Sudarshan}},\ }\bibfield  {title} {\bibinfo {title} {Completely
  positive dynamical semigroups of {$N$}-level systems},\ }\href
  {https://doi.org/10.1063/1.522979} {\bibfield  {journal} {\bibinfo  {journal}
  {J. Math. Phys.}\ }\textbf {\bibinfo {volume} {17}},\ \bibinfo {pages} {821}
  (\bibinfo {year} {1976})}\BibitemShut {NoStop}%
\bibitem [{\citenamefont {Breuer}\ and\ \citenamefont
  {Petruccione}(2007)}]{breuer2007book}%
  \BibitemOpen
  \bibfield  {author} {\bibinfo {author} {\bibfnamefont {H.-P.}\ \bibnamefont
  {Breuer}}\ and\ \bibinfo {author} {\bibfnamefont {F.}~\bibnamefont
  {Petruccione}},\ }\href
  {https://doi.org/10.1093/acprof:oso/9780199213900.001.0001} {\emph {\bibinfo
  {title} {The Theory of Open Quantum Systems}}}\ (\bibinfo  {publisher}
  {Oxford University Press},\ \bibinfo {year} {2007})\BibitemShut {NoStop}%
\bibitem [{\citenamefont {Rivas}\ and\ \citenamefont
  {Huelga}(2012)}]{rivas2012book}%
  \BibitemOpen
  \bibfield  {author} {\bibinfo {author} {\bibfnamefont {{\'A}.}~\bibnamefont
  {Rivas}}\ and\ \bibinfo {author} {\bibfnamefont {S.~F.}\ \bibnamefont
  {Huelga}},\ }\href {https://doi.org/10.1007/978-3-642-23354-8} {\emph
  {\bibinfo {title} {Open Quantum Systems: An Introduction}}}\ (\bibinfo
  {publisher} {Springer Berlin, Heidelberg},\ \bibinfo {year}
  {2012})\BibitemShut {NoStop}%
\bibitem [{\citenamefont {Bardyn}\ \emph {et~al.}(2013)\citenamefont {Bardyn},
  \citenamefont {Baranov}, \citenamefont {V.}, \citenamefont {Rico},
  \citenamefont {İmamoğlu}, \citenamefont {Zoller},\ and\ \citenamefont
  {Diehl}}]{bardyn2013topology}%
  \BibitemOpen
  \bibfield  {author} {\bibinfo {author} {\bibfnamefont {C.-E.}\ \bibnamefont
  {Bardyn}}, \bibinfo {author} {\bibfnamefont {M.~A.}\ \bibnamefont {Baranov}},
  \bibinfo {author} {\bibfnamefont {K.~C.}\ \bibnamefont {V.}}, \bibinfo
  {author} {\bibfnamefont {E.}~\bibnamefont {Rico}}, \bibinfo {author}
  {\bibfnamefont {A.}~\bibnamefont {İmamoğlu}}, \bibinfo {author}
  {\bibfnamefont {P.}~\bibnamefont {Zoller}},\ and\ \bibinfo {author}
  {\bibfnamefont {S.}~\bibnamefont {Diehl}},\ }\bibfield  {title} {\bibinfo
  {title} {Topology by dissipation},\ }\href
  {https://doi.org/10.1088/1367-2630/15/8/085001} {\bibfield  {journal}
  {\bibinfo  {journal} {New J. Phys.}\ }\textbf {\bibinfo {volume} {15}},\
  \bibinfo {pages} {085001} (\bibinfo {year} {2013})}\BibitemShut {NoStop}%
\bibitem [{\citenamefont {Viyuela}\ \emph {et~al.}(2014)\citenamefont
  {Viyuela}, \citenamefont {Rivas},\ and\ \citenamefont
  {Martin-Delgado}}]{viyuela2014uhlmann}%
  \BibitemOpen
  \bibfield  {author} {\bibinfo {author} {\bibfnamefont {O.}~\bibnamefont
  {Viyuela}}, \bibinfo {author} {\bibfnamefont {{\'A}.}~\bibnamefont {Rivas}},\
  and\ \bibinfo {author} {\bibfnamefont {M.~A.}\ \bibnamefont
  {Martin-Delgado}},\ }\bibfield  {title} {\bibinfo {title} {{Uhlmann} phase as
  a topological measure for one-dimensional fermion systems},\ }\href
  {https://doi.org/10.1103/PhysRevLett.112.130401} {\bibfield  {journal}
  {\bibinfo  {journal} {Phys. Rev. Lett.}\ }\textbf {\bibinfo {volume} {112}},\
  \bibinfo {pages} {130401} (\bibinfo {year} {2014})}\BibitemShut {NoStop}%
\bibitem [{\citenamefont {Bardyn}\ \emph {et~al.}(2018)\citenamefont {Bardyn},
  \citenamefont {Wawer}, \citenamefont {Altland}, \citenamefont
  {Fleischhauer},\ and\ \citenamefont {Diehl}}]{bardyn2018probing}%
  \BibitemOpen
  \bibfield  {author} {\bibinfo {author} {\bibfnamefont {C.-E.}\ \bibnamefont
  {Bardyn}}, \bibinfo {author} {\bibfnamefont {L.}~\bibnamefont {Wawer}},
  \bibinfo {author} {\bibfnamefont {A.}~\bibnamefont {Altland}}, \bibinfo
  {author} {\bibfnamefont {M.}~\bibnamefont {Fleischhauer}},\ and\ \bibinfo
  {author} {\bibfnamefont {S.}~\bibnamefont {Diehl}},\ }\bibfield  {title}
  {\bibinfo {title} {Probing the topology of density matrices},\ }\href
  {https://doi.org/10.1103/PhysRevX.8.011035} {\bibfield  {journal} {\bibinfo
  {journal} {Phys. Rev. X}\ }\textbf {\bibinfo {volume} {8}},\ \bibinfo {pages}
  {011035} (\bibinfo {year} {2018})}\BibitemShut {NoStop}%
\bibitem [{\citenamefont {Altland}\ \emph {et~al.}(2021)\citenamefont
  {Altland}, \citenamefont {Fleischhauer},\ and\ \citenamefont
  {Diehl}}]{altland2021symmetry}%
  \BibitemOpen
  \bibfield  {author} {\bibinfo {author} {\bibfnamefont {A.}~\bibnamefont
  {Altland}}, \bibinfo {author} {\bibfnamefont {M.}~\bibnamefont
  {Fleischhauer}},\ and\ \bibinfo {author} {\bibfnamefont {S.}~\bibnamefont
  {Diehl}},\ }\bibfield  {title} {\bibinfo {title} {Symmetry classes of open
  fermionic quantum matter},\ }\href
  {https://doi.org/10.1103/PhysRevX.11.021037} {\bibfield  {journal} {\bibinfo
  {journal} {Phys. Rev. X}\ }\textbf {\bibinfo {volume} {11}},\ \bibinfo
  {pages} {021037} (\bibinfo {year} {2021})}\BibitemShut {NoStop}%
\bibitem [{\citenamefont {van Caspel}\ \emph {et~al.}(2019)\citenamefont {van
  Caspel}, \citenamefont {Tapias~Arze},\ and\ \citenamefont
  {Pérez~Castillo}}]{vancaspel2019dynamical}%
  \BibitemOpen
  \bibfield  {author} {\bibinfo {author} {\bibfnamefont {M.}~\bibnamefont {van
  Caspel}}, \bibinfo {author} {\bibfnamefont {S.~E.}\ \bibnamefont
  {Tapias~Arze}},\ and\ \bibinfo {author} {\bibfnamefont {I.}~\bibnamefont
  {Pérez~Castillo}},\ }\bibfield  {title} {\bibinfo {title} {Dynamical
  signatures of topological order in the driven-dissipative {Kitaev} chain},\
  }\href {https://doi.org/10.21468/SciPostPhys.6.2.026} {\bibfield  {journal}
  {\bibinfo  {journal} {SciPost Phys.}\ }\textbf {\bibinfo {volume} {6}},\
  \bibinfo {pages} {026} (\bibinfo {year} {2019})}\BibitemShut {NoStop}%
\bibitem [{\citenamefont {Lieu}\ \emph {et~al.}(2020)\citenamefont {Lieu},
  \citenamefont {McGinley},\ and\ \citenamefont {Cooper}}]{lieu2020tenfold}%
  \BibitemOpen
  \bibfield  {author} {\bibinfo {author} {\bibfnamefont {S.}~\bibnamefont
  {Lieu}}, \bibinfo {author} {\bibfnamefont {M.}~\bibnamefont {McGinley}},\
  and\ \bibinfo {author} {\bibfnamefont {N.~R.}\ \bibnamefont {Cooper}},\
  }\bibfield  {title} {\bibinfo {title} {Tenfold way for quadratic
  {Lindbladians}},\ }\href {https://doi.org/10.1103/PhysRevLett.124.040401}
  {\bibfield  {journal} {\bibinfo  {journal} {Phys. Rev. Lett.}\ }\textbf
  {\bibinfo {volume} {124}},\ \bibinfo {pages} {040401} (\bibinfo {year}
  {2020})}\BibitemShut {NoStop}%
\bibitem [{\citenamefont {S{\'a}}\ \emph {et~al.}(2023)\citenamefont {S{\'a}},
  \citenamefont {Ribeiro},\ and\ \citenamefont {Prosen}}]{sa2023symmetry}%
  \BibitemOpen
  \bibfield  {author} {\bibinfo {author} {\bibfnamefont {L.}~\bibnamefont
  {S{\'a}}}, \bibinfo {author} {\bibfnamefont {P.}~\bibnamefont {Ribeiro}},\
  and\ \bibinfo {author} {\bibfnamefont {T.}~\bibnamefont {Prosen}},\
  }\bibfield  {title} {\bibinfo {title} {Symmetry classification of many-body
  {Lindbladians}: Tenfold way and beyond},\ }\href
  {https://doi.org/10.1103/PhysRevX.13.031019} {\bibfield  {journal} {\bibinfo
  {journal} {Phys. Rev. X}\ }\textbf {\bibinfo {volume} {13}},\ \bibinfo
  {pages} {031019} (\bibinfo {year} {2023})}\BibitemShut {NoStop}%
\bibitem [{\citenamefont {Mao}\ and\ \citenamefont
  {Yang}(2025)}]{mao2025symmetry}%
  \BibitemOpen
  \bibfield  {author} {\bibinfo {author} {\bibfnamefont {L.}~\bibnamefont
  {Mao}}\ and\ \bibinfo {author} {\bibfnamefont {F.}~\bibnamefont {Yang}},\
  }\bibfield  {title} {\bibinfo {title} {Symmetry classification correspondence
  between quadratic {Lindbladians} and their steady states},\ }\href
  {https://doi.org/10.1103/p8xr-4yzl} {\bibfield  {journal} {\bibinfo
  {journal} {Phys. Rev. B}\ }\textbf {\bibinfo {volume} {112}},\ \bibinfo
  {pages} {085136} (\bibinfo {year} {2025})}\BibitemShut {NoStop}%
\bibitem [{\citenamefont {Chaduteau}\ \emph {et~al.}(2026)\citenamefont
  {Chaduteau}, \citenamefont {Lee},\ and\ \citenamefont
  {Schindler}}]{chaduteau2026postselect}%
  \BibitemOpen
  \bibfield  {author} {\bibinfo {author} {\bibfnamefont {A.}~\bibnamefont
  {Chaduteau}}, \bibinfo {author} {\bibfnamefont {D.~K.~K.}\ \bibnamefont
  {Lee}},\ and\ \bibinfo {author} {\bibfnamefont {F.}~\bibnamefont
  {Schindler}},\ }\bibfield  {title} {\bibinfo {title} {{Lindbladian} versus
  postselected non-{Hermitian} topology},\ }\href
  {https://doi.org/10.1103/ljvt-w6hw} {\bibfield  {journal} {\bibinfo
  {journal} {Phys. Rev. Lett.}\ }\textbf {\bibinfo {volume} {136}},\ \bibinfo
  {pages} {016603} (\bibinfo {year} {2026})}\BibitemShut {NoStop}%
\bibitem [{\citenamefont {Zhu}\ \emph {et~al.}(2021)\citenamefont {Zhu},
  \citenamefont {Teo}, \citenamefont {Li},\ and\ \citenamefont
  {Gong}}]{zhu2021delocalization}%
  \BibitemOpen
  \bibfield  {author} {\bibinfo {author} {\bibfnamefont {W.}~\bibnamefont
  {Zhu}}, \bibinfo {author} {\bibfnamefont {W.~X.}\ \bibnamefont {Teo}},
  \bibinfo {author} {\bibfnamefont {L.}~\bibnamefont {Li}},\ and\ \bibinfo
  {author} {\bibfnamefont {J.}~\bibnamefont {Gong}},\ }\bibfield  {title}
  {\bibinfo {title} {Delocalization of topological edge states},\ }\href
  {https://doi.org/10.1103/PhysRevB.103.195414} {\bibfield  {journal} {\bibinfo
   {journal} {Phys. Rev. B}\ }\textbf {\bibinfo {volume} {103}},\ \bibinfo
  {pages} {195414} (\bibinfo {year} {2021})}\BibitemShut {NoStop}%
\bibitem [{\citenamefont {Wang}\ \emph {et~al.}(2022)\citenamefont {Wang},
  \citenamefont {Wang},\ and\ \citenamefont {Ma}}]{wang2022morphing}%
  \BibitemOpen
  \bibfield  {author} {\bibinfo {author} {\bibfnamefont {W.}~\bibnamefont
  {Wang}}, \bibinfo {author} {\bibfnamefont {X.}~\bibnamefont {Wang}},\ and\
  \bibinfo {author} {\bibfnamefont {G.}~\bibnamefont {Ma}},\ }\bibfield
  {title} {\bibinfo {title} {Non-{Hermitian} morphing of topological modes},\
  }\href {https://doi.org/10.1038/s41586-022-04929-1} {\bibfield  {journal}
  {\bibinfo  {journal} {Nature}\ }\textbf {\bibinfo {volume} {608}},\ \bibinfo
  {pages} {50} (\bibinfo {year} {2022})}\BibitemShut {NoStop}%
\bibitem [{\citenamefont {Yang}\ \emph {et~al.}(2022)\citenamefont {Yang},
  \citenamefont {Jiang},\ and\ \citenamefont {Bergholtz}}]{yang2022skin}%
  \BibitemOpen
  \bibfield  {author} {\bibinfo {author} {\bibfnamefont {F.}~\bibnamefont
  {Yang}}, \bibinfo {author} {\bibfnamefont {Q.-D.}\ \bibnamefont {Jiang}},\
  and\ \bibinfo {author} {\bibfnamefont {E.~J.}\ \bibnamefont {Bergholtz}},\
  }\bibfield  {title} {\bibinfo {title} {{Liouvillian} skin effect in an
  exactly solvable model},\ }\href
  {https://doi.org/10.1103/PhysRevResearch.4.023160} {\bibfield  {journal}
  {\bibinfo  {journal} {Phys. Rev. Res.}\ }\textbf {\bibinfo {volume} {4}},\
  \bibinfo {pages} {023160} (\bibinfo {year} {2022})}\BibitemShut {NoStop}%
\bibitem [{\citenamefont {Zhou}\ and\ \citenamefont
  {Yu}(2022)}]{zhou2022adjoint}%
  \BibitemOpen
  \bibfield  {author} {\bibinfo {author} {\bibfnamefont {Z.}~\bibnamefont
  {Zhou}}\ and\ \bibinfo {author} {\bibfnamefont {Z.}~\bibnamefont {Yu}},\
  }\bibfield  {title} {\bibinfo {title} {Non-{Hermitian} skin effect in
  quadratic {Lindbladian} systems: An adjoint fermion approach},\ }\href
  {https://doi.org/10.1103/PhysRevA.106.032216} {\bibfield  {journal} {\bibinfo
   {journal} {Phys. Rev. A}\ }\textbf {\bibinfo {volume} {106}},\ \bibinfo
  {pages} {032216} (\bibinfo {year} {2022})}\BibitemShut {NoStop}%
\bibitem [{\citenamefont {Lin}\ \emph {et~al.}(2023)\citenamefont {Lin},
  \citenamefont {Tai}, \citenamefont {Li},\ and\ \citenamefont
  {Lee}}]{lin2023skin}%
  \BibitemOpen
  \bibfield  {author} {\bibinfo {author} {\bibfnamefont {R.}~\bibnamefont
  {Lin}}, \bibinfo {author} {\bibfnamefont {T.}~\bibnamefont {Tai}}, \bibinfo
  {author} {\bibfnamefont {L.}~\bibnamefont {Li}},\ and\ \bibinfo {author}
  {\bibfnamefont {C.-H.}\ \bibnamefont {Lee}},\ }\bibfield  {title} {\bibinfo
  {title} {Topological non-{Hermitian} skin effect},\ }\href
  {https://doi.org/10.1007/s11467-023-1309-z} {\bibfield  {journal} {\bibinfo
  {journal} {Front. Phys.}\ }\textbf {\bibinfo {volume} {18}},\ \bibinfo
  {pages} {53605} (\bibinfo {year} {2023})}\BibitemShut {NoStop}%
\bibitem [{\citenamefont {Haga}\ \emph {et~al.}(2021)\citenamefont {Haga},
  \citenamefont {Nakagawa}, \citenamefont {Hamazaki},\ and\ \citenamefont
  {Ueda}}]{haga2021slowing}%
  \BibitemOpen
  \bibfield  {author} {\bibinfo {author} {\bibfnamefont {T.}~\bibnamefont
  {Haga}}, \bibinfo {author} {\bibfnamefont {M.}~\bibnamefont {Nakagawa}},
  \bibinfo {author} {\bibfnamefont {R.}~\bibnamefont {Hamazaki}},\ and\
  \bibinfo {author} {\bibfnamefont {M.}~\bibnamefont {Ueda}},\ }\bibfield
  {title} {\bibinfo {title} {{Liouvillian} skin effect: Slowing down of
  relaxation processes without gap closing},\ }\href
  {https://doi.org/10.1103/PhysRevLett.127.070402} {\bibfield  {journal}
  {\bibinfo  {journal} {Phys. Rev. Lett.}\ }\textbf {\bibinfo {volume} {127}},\
  \bibinfo {pages} {070402} (\bibinfo {year} {2021})}\BibitemShut {NoStop}%
\bibitem [{\citenamefont {Song}\ \emph {et~al.}(2019)\citenamefont {Song},
  \citenamefont {Yao},\ and\ \citenamefont {Wang}}]{song2019chiral}%
  \BibitemOpen
  \bibfield  {author} {\bibinfo {author} {\bibfnamefont {F.}~\bibnamefont
  {Song}}, \bibinfo {author} {\bibfnamefont {S.}~\bibnamefont {Yao}},\ and\
  \bibinfo {author} {\bibfnamefont {Z.}~\bibnamefont {Wang}},\ }\bibfield
  {title} {\bibinfo {title} {Non-{Hermitian} skin effect and chiral damping in
  open quantum systems},\ }\href
  {https://doi.org/10.1103/PhysRevLett.123.170401} {\bibfield  {journal}
  {\bibinfo  {journal} {Phys. Rev. Lett.}\ }\textbf {\bibinfo {volume} {123}},\
  \bibinfo {pages} {170401} (\bibinfo {year} {2019})}\BibitemShut {NoStop}%
\bibitem [{\citenamefont {Sarkar}\ \emph {et~al.}(2026)\citenamefont {Sarkar},
  \citenamefont {Hegde}, \citenamefont {Narayan},\ and\ \citenamefont
  {Meng}}]{sarkar2026chiral}%
  \BibitemOpen
  \bibfield  {author} {\bibinfo {author} {\bibfnamefont {R.}~\bibnamefont
  {Sarkar}}, \bibinfo {author} {\bibfnamefont {S.~S.}\ \bibnamefont {Hegde}},
  \bibinfo {author} {\bibfnamefont {A.}~\bibnamefont {Narayan}},\ and\ \bibinfo
  {author} {\bibfnamefont {T.}~\bibnamefont {Meng}},\ }\bibfield  {title}
  {\bibinfo {title} {Chiral damping with persistent edge states from the
  interplay of topologies in open quantum systems},\ }\href
  {https://doi.org/10.1038/s42005-026-02573-z} {\bibfield  {journal} {\bibinfo
  {journal} {Commun. Phys.}\ }\textbf {\bibinfo {volume} {9}},\ \bibinfo
  {pages} {109} (\bibinfo {year} {2026})}\BibitemShut {NoStop}%
\bibitem [{\citenamefont {Yang}\ \emph {et~al.}(2026)\citenamefont {Yang},
  \citenamefont {Zelenayova}, \citenamefont {Molignini},\ and\ \citenamefont
  {Bergholtz}}]{yang2026boundary}%
  \BibitemOpen
  \bibfield  {author} {\bibinfo {author} {\bibfnamefont {F.}~\bibnamefont
  {Yang}}, \bibinfo {author} {\bibfnamefont {M.}~\bibnamefont {Zelenayova}},
  \bibinfo {author} {\bibfnamefont {P.}~\bibnamefont {Molignini}},\ and\
  \bibinfo {author} {\bibfnamefont {E.~J.}\ \bibnamefont {Bergholtz}},\ }\href
  {https://arxiv.org/abs/2506.16308} {\bibinfo {title} {Quantum dynamical
  signatures of non-{Hermitian} boundary modes}} (\bibinfo {year} {2026}),\
  \Eprint {https://arxiv.org/abs/2506.16308} {arXiv:2506.16308
  [cond-mat.mes-hall]} \BibitemShut {NoStop}%
\bibitem [{\citenamefont {Sayyad}\ \emph {et~al.}(2021)\citenamefont {Sayyad},
  \citenamefont {Yu}, \citenamefont {Grushin},\ and\ \citenamefont
  {Sieberer}}]{sayyad2021entanglement}%
  \BibitemOpen
  \bibfield  {author} {\bibinfo {author} {\bibfnamefont {S.}~\bibnamefont
  {Sayyad}}, \bibinfo {author} {\bibfnamefont {J.}~\bibnamefont {Yu}}, \bibinfo
  {author} {\bibfnamefont {A.~G.}\ \bibnamefont {Grushin}},\ and\ \bibinfo
  {author} {\bibfnamefont {L.~M.}\ \bibnamefont {Sieberer}},\ }\bibfield
  {title} {\bibinfo {title} {Entanglement spectrum crossings reveal
  non-{Hermitian} dynamical topology},\ }\href
  {https://doi.org/10.1103/PhysRevResearch.3.033022} {\bibfield  {journal}
  {\bibinfo  {journal} {Phys. Rev. Res.}\ }\textbf {\bibinfo {volume} {3}},\
  \bibinfo {pages} {033022} (\bibinfo {year} {2021})}\BibitemShut {NoStop}%
\bibitem [{\citenamefont {Bayona-Pena}\ \emph {et~al.}(2025)\citenamefont
  {Bayona-Pena}, \citenamefont {Hanai}, \citenamefont {Mori},\ and\
  \citenamefont {Hayakawa}}]{bayona2025entanglement}%
  \BibitemOpen
  \bibfield  {author} {\bibinfo {author} {\bibfnamefont {P.}~\bibnamefont
  {Bayona-Pena}}, \bibinfo {author} {\bibfnamefont {R.}~\bibnamefont {Hanai}},
  \bibinfo {author} {\bibfnamefont {T.}~\bibnamefont {Mori}},\ and\ \bibinfo
  {author} {\bibfnamefont {H.}~\bibnamefont {Hayakawa}},\ }\bibfield  {title}
  {\bibinfo {title} {Entanglement spectrum dynamics as a probe for
  non-{Hermitian} bulk-boundary correspondence in systems with periodic
  boundaries},\ }\href {https://doi.org/10.1103/PhysRevB.111.L140303}
  {\bibfield  {journal} {\bibinfo  {journal} {Phys. Rev. B}\ }\textbf {\bibinfo
  {volume} {111}},\ \bibinfo {pages} {L140303} (\bibinfo {year}
  {2025})}\BibitemShut {NoStop}%
\bibitem [{\citenamefont {Wang}\ \emph {et~al.}(2021)\citenamefont {Wang},
  \citenamefont {Li}, \citenamefont {Xiao}, \citenamefont {Han}, \citenamefont
  {Yi},\ and\ \citenamefont {Xue}}]{wang2021detecting}%
  \BibitemOpen
  \bibfield  {author} {\bibinfo {author} {\bibfnamefont {K.}~\bibnamefont
  {Wang}}, \bibinfo {author} {\bibfnamefont {T.}~\bibnamefont {Li}}, \bibinfo
  {author} {\bibfnamefont {L.}~\bibnamefont {Xiao}}, \bibinfo {author}
  {\bibfnamefont {Y.}~\bibnamefont {Han}}, \bibinfo {author} {\bibfnamefont
  {W.}~\bibnamefont {Yi}},\ and\ \bibinfo {author} {\bibfnamefont
  {P.}~\bibnamefont {Xue}},\ }\bibfield  {title} {\bibinfo {title} {Detecting
  non-{Bloch} topological invariants in quantum dynamics},\ }\href
  {https://doi.org/10.1103/PhysRevLett.127.270602} {\bibfield  {journal}
  {\bibinfo  {journal} {Phys. Rev. Lett.}\ }\textbf {\bibinfo {volume} {127}},\
  \bibinfo {pages} {270602} (\bibinfo {year} {2021})}\BibitemShut {NoStop}%
\bibitem [{\citenamefont {Lin}\ and\ \citenamefont
  {Zhang}(2025)}]{lin2025quench}%
  \BibitemOpen
  \bibfield  {author} {\bibinfo {author} {\bibfnamefont {X.-D.}\ \bibnamefont
  {Lin}}\ and\ \bibinfo {author} {\bibfnamefont {L.}~\bibnamefont {Zhang}},\
  }\bibfield  {title} {\bibinfo {title} {Measuring non-{Hermitian} topological
  invariants directly from quench dynamics},\ }\href
  {https://doi.org/10.1103/PhysRevResearch.7.L012060} {\bibfield  {journal}
  {\bibinfo  {journal} {Phys. Rev. Res.}\ }\textbf {\bibinfo {volume} {7}},\
  \bibinfo {pages} {L012060} (\bibinfo {year} {2025})}\BibitemShut {NoStop}%
\bibitem [{\citenamefont {Kitaev}(2001)}]{kitaev2001original}%
  \BibitemOpen
  \bibfield  {author} {\bibinfo {author} {\bibfnamefont {A.~Y.}\ \bibnamefont
  {Kitaev}},\ }\bibfield  {title} {\bibinfo {title} {Unpaired {Majorana}
  fermions in quantum wires},\ }\href
  {https://doi.org/10.1070/1063-7869/44/10S/S29} {\bibfield  {journal}
  {\bibinfo  {journal} {Phys. Usp.}\ }\textbf {\bibinfo {volume} {44}},\
  \bibinfo {pages} {131} (\bibinfo {year} {2001})}\BibitemShut {NoStop}%
\bibitem [{\citenamefont {Metelmann}\ and\ \citenamefont
  {Clerk}(2015)}]{metelmann2015nonreciprocal}%
  \BibitemOpen
  \bibfield  {author} {\bibinfo {author} {\bibfnamefont {A.}~\bibnamefont
  {Metelmann}}\ and\ \bibinfo {author} {\bibfnamefont {A.~A.}\ \bibnamefont
  {Clerk}},\ }\bibfield  {title} {\bibinfo {title} {Nonreciprocal photon
  transmission and amplification via reservoir engineering},\ }\href
  {https://doi.org/10.1103/PhysRevX.5.021025} {\bibfield  {journal} {\bibinfo
  {journal} {Phys. Rev. X}\ }\textbf {\bibinfo {volume} {5}},\ \bibinfo {pages}
  {021025} (\bibinfo {year} {2015})}\BibitemShut {NoStop}%
\bibitem [{\citenamefont {McDonald}\ \emph {et~al.}(2022)\citenamefont
  {McDonald}, \citenamefont {Hanai},\ and\ \citenamefont
  {Clerk}}]{mcdonald2022nonreciprocal}%
  \BibitemOpen
  \bibfield  {author} {\bibinfo {author} {\bibfnamefont {A.}~\bibnamefont
  {McDonald}}, \bibinfo {author} {\bibfnamefont {R.}~\bibnamefont {Hanai}},\
  and\ \bibinfo {author} {\bibfnamefont {A.~A.}\ \bibnamefont {Clerk}},\
  }\bibfield  {title} {\bibinfo {title} {Nonequilibrium stationary states of
  quantum non-{Hermitian} lattice models},\ }\href
  {https://doi.org/10.1103/PhysRevB.105.064302} {\bibfield  {journal} {\bibinfo
   {journal} {Phys. Rev. B}\ }\textbf {\bibinfo {volume} {105}},\ \bibinfo
  {pages} {064302} (\bibinfo {year} {2022})}\BibitemShut {NoStop}%
\bibitem [{\citenamefont {Soares}\ \emph {et~al.}(2026)\citenamefont {Soares},
  \citenamefont {Brunelli},\ and\ \citenamefont
  {Schir{\`o}}}]{soares2026nonreciprocal}%
  \BibitemOpen
  \bibfield  {author} {\bibinfo {author} {\bibfnamefont {R.~D.}\ \bibnamefont
  {Soares}}, \bibinfo {author} {\bibfnamefont {M.}~\bibnamefont {Brunelli}},\
  and\ \bibinfo {author} {\bibfnamefont {M.}~\bibnamefont {Schir{\`o}}},\
  }\bibfield  {title} {\bibinfo {title} {Dissipative phase transition of
  interacting nonreciprocal fermions},\ }\href
  {https://doi.org/10.1103/n8xv-kygx} {\bibfield  {journal} {\bibinfo
  {journal} {Phys. Rev. Lett.}\ }\textbf {\bibinfo {volume} {136}},\ \bibinfo
  {pages} {250403} (\bibinfo {year} {2026})}\BibitemShut {NoStop}%
\bibitem [{\citenamefont {Brighi}\ and\ \citenamefont
  {Nunnenkamp}(2026)}]{brighi2026kitaev}%
  \BibitemOpen
  \bibfield  {author} {\bibinfo {author} {\bibfnamefont {P.}~\bibnamefont
  {Brighi}}\ and\ \bibinfo {author} {\bibfnamefont {A.}~\bibnamefont
  {Nunnenkamp}},\ }\bibfield  {title} {\bibinfo {title} {Pairing induced phase
  transition in the non reciprocal {Kitaev} chain},\ }\href
  {https://doi.org/10.1038/s42005-026-02752-y} {\bibfield  {journal} {\bibinfo
  {journal} {Commun. Phys.}\ } (\bibinfo {year} {2026})}\BibitemShut {NoStop}%
\bibitem [{\citenamefont {D'Abbruzzo}\ and\ \citenamefont
  {Rossini}(2021{\natexlab{a}})}]{dabbruzzo2021self}%
  \BibitemOpen
  \bibfield  {author} {\bibinfo {author} {\bibfnamefont {A.}~\bibnamefont
  {D'Abbruzzo}}\ and\ \bibinfo {author} {\bibfnamefont {D.}~\bibnamefont
  {Rossini}},\ }\bibfield  {title} {\bibinfo {title} {Self-consistent
  microscopic derivation of {Markovian} master equations for open quadratic
  quantum systems},\ }\href {https://doi.org/10.1103/PhysRevA.103.052209}
  {\bibfield  {journal} {\bibinfo  {journal} {Phys. Rev. A}\ }\textbf {\bibinfo
  {volume} {103}},\ \bibinfo {pages} {052209} (\bibinfo {year}
  {2021}{\natexlab{a}})}\BibitemShut {NoStop}%
\bibitem [{\citenamefont {Prosen}(2008)}]{prosen2008third}%
  \BibitemOpen
  \bibfield  {author} {\bibinfo {author} {\bibfnamefont {T.}~\bibnamefont
  {Prosen}},\ }\bibfield  {title} {\bibinfo {title} {Third quantization: A
  general method to solve master equations for quadratic open {Fermi}
  systems},\ }\href {https://doi.org/10.1088/1367-2630/10/4/043026} {\bibfield
  {journal} {\bibinfo  {journal} {New J. Phys.}\ }\textbf {\bibinfo {volume}
  {10}},\ \bibinfo {pages} {043026} (\bibinfo {year} {2008})}\BibitemShut
  {NoStop}%
\bibitem [{\citenamefont {Barthel}\ and\ \citenamefont
  {Zhang}(2022)}]{barthel2022solving}%
  \BibitemOpen
  \bibfield  {author} {\bibinfo {author} {\bibfnamefont {T.}~\bibnamefont
  {Barthel}}\ and\ \bibinfo {author} {\bibfnamefont {Y.}~\bibnamefont
  {Zhang}},\ }\bibfield  {title} {\bibinfo {title} {Solving quasi-free and
  quadratic {Lindblad} master equations for open fermionic and bosonic
  systems},\ }\href {https://doi.org/10.1088/1742-5468/ac8e5c} {\bibfield
  {journal} {\bibinfo  {journal} {J. Stat. Mech.: Theor. Exp.}\ }\textbf
  {\bibinfo {volume} {2022}},\ \bibinfo {pages} {113101} (\bibinfo {year}
  {2022})}\BibitemShut {NoStop}%
\bibitem [{\citenamefont {Nigro}(2019)}]{nigro2019uniqueness}%
  \BibitemOpen
  \bibfield  {author} {\bibinfo {author} {\bibfnamefont {D.}~\bibnamefont
  {Nigro}},\ }\bibfield  {title} {\bibinfo {title} {On the uniqueness of the
  steady-state solution of the {Lindblad–Gorini–Kossakowski–Sudarshan}
  equation},\ }\href {https://doi.org/10.1088/1742-5468/ab0c1c} {\bibfield
  {journal} {\bibinfo  {journal} {J. Stat. Mech.: Theory Exp.}\ }\textbf
  {\bibinfo {volume} {2019}}\bibinfo  {number} { (4)},\ \bibinfo {pages}
  {043202}}\BibitemShut {NoStop}%
\bibitem [{\citenamefont {Bernard}\ and\ \citenamefont
  {LeClair}(2002)}]{bernard2002classification}%
  \BibitemOpen
\bibfield  {number} {  }\bibfield  {author} {\bibinfo {author} {\bibfnamefont
  {D.}~\bibnamefont {Bernard}}\ and\ \bibinfo {author} {\bibfnamefont
  {A.}~\bibnamefont {LeClair}},\ }\bibinfo {title} {A classification of
  non-{Hermitian} random matrices},\ in\ \href
  {https://doi.org/10.1007/978-94-010-0514-2_19} {\emph {\bibinfo {booktitle}
  {Statistical Field Theories}}},\ \bibinfo {editor} {edited by\ \bibinfo
  {editor} {\bibfnamefont {A.}~\bibnamefont {Cappelli}}\ and\ \bibinfo {editor}
  {\bibfnamefont {G.}~\bibnamefont {Mussardo}}}\ (\bibinfo  {publisher}
  {Springer Netherlands},\ \bibinfo {year} {2002})\ pp.\ \bibinfo {pages}
  {207--214}\BibitemShut {NoStop}%
\bibitem [{\citenamefont {Kawabata}\ \emph {et~al.}(2019)\citenamefont
  {Kawabata}, \citenamefont {Shiozaki}, \citenamefont {Ueda},\ and\
  \citenamefont {Sato}}]{kawabata2019symmetry}%
  \BibitemOpen
  \bibfield  {author} {\bibinfo {author} {\bibfnamefont {K.}~\bibnamefont
  {Kawabata}}, \bibinfo {author} {\bibfnamefont {K.}~\bibnamefont {Shiozaki}},
  \bibinfo {author} {\bibfnamefont {M.}~\bibnamefont {Ueda}},\ and\ \bibinfo
  {author} {\bibfnamefont {M.}~\bibnamefont {Sato}},\ }\bibfield  {title}
  {\bibinfo {title} {Symmetry and topology in non-{Hermitian} physics},\ }\href
  {https://doi.org/10.1103/PhysRevX.9.041015} {\bibfield  {journal} {\bibinfo
  {journal} {Phys. Rev. X}\ }\textbf {\bibinfo {volume} {9}},\ \bibinfo {pages}
  {041015} (\bibinfo {year} {2019})}\BibitemShut {NoStop}%
\bibitem [{\citenamefont {Kawasaki}\ \emph {et~al.}(2022)\citenamefont
  {Kawasaki}, \citenamefont {Mochizuki},\ and\ \citenamefont
  {Obuse}}]{kawasaki2022shifted}%
  \BibitemOpen
  \bibfield  {author} {\bibinfo {author} {\bibfnamefont {M.}~\bibnamefont
  {Kawasaki}}, \bibinfo {author} {\bibfnamefont {K.}~\bibnamefont
  {Mochizuki}},\ and\ \bibinfo {author} {\bibfnamefont {H.}~\bibnamefont
  {Obuse}},\ }\bibfield  {title} {\bibinfo {title} {Topological phases
  protected by shifted sublattice symmetry in dissipative quantum systems},\
  }\href {https://doi.org/10.1103/PhysRevB.106.035408} {\bibfield  {journal}
  {\bibinfo  {journal} {Phys. Rev. B}\ }\textbf {\bibinfo {volume} {106}},\
  \bibinfo {pages} {035408} (\bibinfo {year} {2022})}\BibitemShut {NoStop}%
\bibitem [{\citenamefont {Böttcher}\ and\ \citenamefont
  {Silbermann}(1999)}]{bottcher1999book}%
  \BibitemOpen
  \bibfield  {author} {\bibinfo {author} {\bibfnamefont {A.}~\bibnamefont
  {Böttcher}}\ and\ \bibinfo {author} {\bibfnamefont {B.}~\bibnamefont
  {Silbermann}},\ }\href {https://doi.org/10.1007/978-1-4612-1426-7} {\emph
  {\bibinfo {title} {Introduction to Large Truncated {Toeplitz} Matrices}}}\
  (\bibinfo  {publisher} {Springer New York},\ \bibinfo {year}
  {1999})\BibitemShut {NoStop}%
\bibitem [{\citenamefont {Yokomizo}\ and\ \citenamefont
  {Murakami}(2019)}]{yokomizo2019gbz}%
  \BibitemOpen
  \bibfield  {author} {\bibinfo {author} {\bibfnamefont {K.}~\bibnamefont
  {Yokomizo}}\ and\ \bibinfo {author} {\bibfnamefont {S.}~\bibnamefont
  {Murakami}},\ }\bibfield  {title} {\bibinfo {title} {Non-{Bloch} band theory
  of non-{Hermitian} systems},\ }\href
  {https://doi.org/10.1103/PhysRevLett.123.066404} {\bibfield  {journal}
  {\bibinfo  {journal} {Phys. Rev. Lett.}\ }\textbf {\bibinfo {volume} {123}},\
  \bibinfo {pages} {066404} (\bibinfo {year} {2019})}\BibitemShut {NoStop}%
\bibitem [{\citenamefont {Yang}\ \emph {et~al.}(2020)\citenamefont {Yang},
  \citenamefont {Zhang}, \citenamefont {Fang},\ and\ \citenamefont
  {Hu}}]{yang2020auxiliary}%
  \BibitemOpen
  \bibfield  {author} {\bibinfo {author} {\bibfnamefont {Z.}~\bibnamefont
  {Yang}}, \bibinfo {author} {\bibfnamefont {K.}~\bibnamefont {Zhang}},
  \bibinfo {author} {\bibfnamefont {C.}~\bibnamefont {Fang}},\ and\ \bibinfo
  {author} {\bibfnamefont {J.}~\bibnamefont {Hu}},\ }\bibfield  {title}
  {\bibinfo {title} {Non-{Hermitian} bulk-boundary correspondence and auxiliary
  generalized {Brillouin} zone theory},\ }\href
  {https://doi.org/10.1103/PhysRevLett.125.226402} {\bibfield  {journal}
  {\bibinfo  {journal} {Phys. Rev. Lett.}\ }\textbf {\bibinfo {volume} {125}},\
  \bibinfo {pages} {226402} (\bibinfo {year} {2020})}\BibitemShut {NoStop}%
\bibitem [{\citenamefont {Wang}\ \emph {et~al.}(2026)\citenamefont {Wang},
  \citenamefont {Zhong},\ and\ \citenamefont {Fan}}]{wang2026gbz}%
  \BibitemOpen
  \bibfield  {author} {\bibinfo {author} {\bibfnamefont {H.}~\bibnamefont
  {Wang}}, \bibinfo {author} {\bibfnamefont {J.}~\bibnamefont {Zhong}},\ and\
  \bibinfo {author} {\bibfnamefont {S.}~\bibnamefont {Fan}},\ }\bibfield
  {title} {\bibinfo {title} {Topology of the generalized {Brillouin} zone of
  one-dimensional models},\ }\href {https://doi.org/10.1103/7cnd-r7q4}
  {\bibfield  {journal} {\bibinfo  {journal} {Phys. Rev. Lett.}\ }\textbf
  {\bibinfo {volume} {136}},\ \bibinfo {pages} {206601} (\bibinfo {year}
  {2026})}\BibitemShut {NoStop}%
\bibitem [{sm()}]{sm}%
  \BibitemOpen
  \href@noop {} {}\bibinfo {note} {See Supplemental Material for derivations
  regarding the finite-size formulation of the protocol and the phase diagram
  of the nonreciprocal Kitaev chain.}\BibitemShut {Stop}%
\bibitem [{\citenamefont {Trefethen}\ and\ \citenamefont
  {Embree}(2005)}]{trefethen2005book}%
  \BibitemOpen
  \bibfield  {author} {\bibinfo {author} {\bibfnamefont {L.~N.}\ \bibnamefont
  {Trefethen}}\ and\ \bibinfo {author} {\bibfnamefont {M.}~\bibnamefont
  {Embree}},\ }\href {https://doi.org/10.1515/9780691213101} {\emph {\bibinfo
  {title} {Spectra and Pseudospectra: The Behavior of Nonnormal Matrices and
  Operators}}}\ (\bibinfo  {publisher} {Princeton University Press},\ \bibinfo
  {year} {2005})\BibitemShut {NoStop}%
\bibitem [{\citenamefont {Nakatsukasa}\ \emph {et~al.}(2018)\citenamefont
  {Nakatsukasa}, \citenamefont {S{\`e}te},\ and\ \citenamefont
  {Trefethen}}]{nakatsukasa2018aaa}%
  \BibitemOpen
  \bibfield  {author} {\bibinfo {author} {\bibfnamefont {Y.}~\bibnamefont
  {Nakatsukasa}}, \bibinfo {author} {\bibfnamefont {O.}~\bibnamefont
  {S{\`e}te}},\ and\ \bibinfo {author} {\bibfnamefont {L.~N.}\ \bibnamefont
  {Trefethen}},\ }\bibfield  {title} {\bibinfo {title} {The {AAA} algorithm for
  rational approximation},\ }\href {https://doi.org/10.1137/16M1106122}
  {\bibfield  {journal} {\bibinfo  {journal} {SIAM J. Sci. Comput.}\ }\textbf
  {\bibinfo {volume} {40}},\ \bibinfo {pages} {A1494} (\bibinfo {year}
  {2018})}\BibitemShut {NoStop}%
\bibitem [{\citenamefont {Lee}\ and\ \citenamefont
  {Thomale}(2019)}]{lee2019anatomy}%
  \BibitemOpen
  \bibfield  {author} {\bibinfo {author} {\bibfnamefont {C.~H.}\ \bibnamefont
  {Lee}}\ and\ \bibinfo {author} {\bibfnamefont {R.}~\bibnamefont {Thomale}},\
  }\bibfield  {title} {\bibinfo {title} {Anatomy of skin modes and topology in
  non-{Hermitian} systems},\ }\href
  {https://doi.org/10.1103/PhysRevB.99.201103} {\bibfield  {journal} {\bibinfo
  {journal} {Phys. Rev. B}\ }\textbf {\bibinfo {volume} {99}},\ \bibinfo
  {pages} {201103(R)} (\bibinfo {year} {2019})}\BibitemShut {NoStop}%
\bibitem [{\citenamefont {Zhong}\ \emph {et~al.}(2024)\citenamefont {Zhong},
  \citenamefont {Wang},\ and\ \citenamefont {Fan}}]{zhong2024pole}%
  \BibitemOpen
  \bibfield  {author} {\bibinfo {author} {\bibfnamefont {J.}~\bibnamefont
  {Zhong}}, \bibinfo {author} {\bibfnamefont {H.}~\bibnamefont {Wang}},\ and\
  \bibinfo {author} {\bibfnamefont {S.}~\bibnamefont {Fan}},\ }\bibfield
  {title} {\bibinfo {title} {Pole and zero edge state invariant for
  one-dimensional non-{Hermitian} sublattice symmetry},\ }\href
  {https://doi.org/10.1103/PhysRevB.110.214113} {\bibfield  {journal} {\bibinfo
   {journal} {Phys. Rev. B}\ }\textbf {\bibinfo {volume} {110}},\ \bibinfo
  {pages} {214113} (\bibinfo {year} {2024})}\BibitemShut {NoStop}%
\bibitem [{\citenamefont {Sveshnikov}\ and\ \citenamefont
  {Tikhonov}(1978)}]{sveshnikov1978book}%
  \BibitemOpen
  \bibfield  {author} {\bibinfo {author} {\bibfnamefont {A.~G.}\ \bibnamefont
  {Sveshnikov}}\ and\ \bibinfo {author} {\bibfnamefont {A.~N.}\ \bibnamefont
  {Tikhonov}},\ }\href@noop {} {\emph {\bibinfo {title} {The Theory of
  Functions of a Complex Variable}}},\ \bibinfo {edition} {2nd}\ ed.\ (\bibinfo
   {publisher} {Mir Publishers},\ \bibinfo {year} {1978})\BibitemShut {NoStop}%
\bibitem [{\citenamefont {Zhao}\ \emph {et~al.}(2021)\citenamefont {Zhao},
  \citenamefont {Guo}, \citenamefont {Kou}, \citenamefont {Zhuang},\ and\
  \citenamefont {Liu}}]{zhao2021defective}%
  \BibitemOpen
  \bibfield  {author} {\bibinfo {author} {\bibfnamefont {X.-M.}\ \bibnamefont
  {Zhao}}, \bibinfo {author} {\bibfnamefont {C.-X.}\ \bibnamefont {Guo}},
  \bibinfo {author} {\bibfnamefont {S.-P.}\ \bibnamefont {Kou}}, \bibinfo
  {author} {\bibfnamefont {L.}~\bibnamefont {Zhuang}},\ and\ \bibinfo {author}
  {\bibfnamefont {W.-M.}\ \bibnamefont {Liu}},\ }\bibfield  {title} {\bibinfo
  {title} {Defective {Majorana} zero modes in a non-{Hermitian} {Kitaev}
  chain},\ }\href {https://doi.org/10.1103/PhysRevB.104.205131} {\bibfield
  {journal} {\bibinfo  {journal} {Phys. Rev. B}\ }\textbf {\bibinfo {volume}
  {104}},\ \bibinfo {pages} {205131} (\bibinfo {year} {2021})}\BibitemShut
  {NoStop}%
\bibitem [{\citenamefont {Ardonne}\ and\ \citenamefont
  {Kurasov}(2025)}]{ardonne2025kitaev}%
  \BibitemOpen
  \bibfield  {author} {\bibinfo {author} {\bibfnamefont {E.}~\bibnamefont
  {Ardonne}}\ and\ \bibinfo {author} {\bibfnamefont {V.}~\bibnamefont
  {Kurasov}},\ }\bibfield  {title} {\bibinfo {title} {On the non-hermitian
  {Kitaev} chain},\ }\href {https://doi.org/10.1088/1751-8121/add4d3}
  {\bibfield  {journal} {\bibinfo  {journal} {J. Phys. A: Math. Theor.}\
  }\textbf {\bibinfo {volume} {58}},\ \bibinfo {pages} {205302} (\bibinfo
  {year} {2025})}\BibitemShut {NoStop}%
\bibitem [{\citenamefont {Raj}\ \emph {et~al.}(2026)\citenamefont {Raj},
  \citenamefont {Ray},\ and\ \citenamefont {Samal}}]{ray2026kitaev}%
  \BibitemOpen
  \bibfield  {author} {\bibinfo {author} {\bibfnamefont {A.}~\bibnamefont
  {Raj}}, \bibinfo {author} {\bibfnamefont {S.}~\bibnamefont {Ray}},\ and\
  \bibinfo {author} {\bibfnamefont {S.~S.}\ \bibnamefont {Samal}},\ }\bibfield
  {title} {\bibinfo {title} {Symmetry and topology in a non-{Hermitian}
  {Kitaev} chain},\ }\href {https://doi.org/10.1103/mhwj-5c85} {\bibfield
  {journal} {\bibinfo  {journal} {Phys. Rev. B}\ }\textbf {\bibinfo {volume}
  {113}},\ \bibinfo {pages} {235104} (\bibinfo {year} {2026})}\BibitemShut
  {NoStop}%
\bibitem [{\citenamefont {D'Abbruzzo}\ and\ \citenamefont
  {Rossini}(2021{\natexlab{b}})}]{dabbruzzo2021kitaev}%
  \BibitemOpen
  \bibfield  {author} {\bibinfo {author} {\bibfnamefont {A.}~\bibnamefont
  {D'Abbruzzo}}\ and\ \bibinfo {author} {\bibfnamefont {D.}~\bibnamefont
  {Rossini}},\ }\bibfield  {title} {\bibinfo {title} {Topological signatures in
  a weakly dissipative {Kitaev} chain of finite length},\ }\href
  {https://doi.org/10.1103/PhysRevB.104.115139} {\bibfield  {journal} {\bibinfo
   {journal} {Phys. Rev. B}\ }\textbf {\bibinfo {volume} {104}},\ \bibinfo
  {pages} {115139} (\bibinfo {year} {2021}{\natexlab{b}})}\BibitemShut
  {NoStop}%
\end{thebibliography}%


\begin{thebibliography}{6}%
\makeatletter
\providecommand \@ifxundefined [1]{%
 \@ifx{#1\undefined}
}%
\providecommand \@ifnum [1]{%
 \ifnum #1\expandafter \@firstoftwo
 \else \expandafter \@secondoftwo
 \fi
}%
\providecommand \@ifx [1]{%
 \ifx #1\expandafter \@firstoftwo
 \else \expandafter \@secondoftwo
 \fi
}%
\providecommand \natexlab [1]{#1}%
\providecommand \enquote  [1]{``#1''}%
\providecommand \bibnamefont  [1]{#1}%
\providecommand \bibfnamefont [1]{#1}%
\providecommand \citenamefont [1]{#1}%
\providecommand \href@noop [0]{\@secondoftwo}%
\providecommand \href [0]{\begingroup \@sanitize@url \@href}%
\providecommand \@href[1]{\@@startlink{#1}\@@href}%
\providecommand \@@href[1]{\endgroup#1\@@endlink}%
\providecommand \@sanitize@url [0]{\catcode `\\12\catcode `\$12\catcode
  `\&12\catcode `\#12\catcode `\^12\catcode `\_12\catcode `\%12\relax}%
\providecommand \@@startlink[1]{}%
\providecommand \@@endlink[0]{}%
\providecommand \url  [0]{\begingroup\@sanitize@url \@url }%
\providecommand \@url [1]{\endgroup\@href {#1}{\urlprefix }}%
\providecommand \urlprefix  [0]{URL }%
\providecommand \Eprint [0]{\href }%
\providecommand \doibase [0]{https://doi.org/}%
\providecommand \selectlanguage [0]{\@gobble}%
\providecommand \bibinfo  [0]{\@secondoftwo}%
\providecommand \bibfield  [0]{\@secondoftwo}%
\providecommand \translation [1]{[#1]}%
\providecommand \BibitemOpen [0]{}%
\providecommand \bibitemStop [0]{}%
\providecommand \bibitemNoStop [0]{.\EOS\space}%
\providecommand \EOS [0]{\spacefactor3000\relax}%
\providecommand \BibitemShut  [1]{\csname bibitem#1\endcsname}%
\let\auto@bib@innerbib\@empty
\bibitem [{\citenamefont {Böttcher}\ and\ \citenamefont
  {Silbermann}(1999)}]{bottcher1999book}%
  \BibitemOpen
  \bibfield  {author} {\bibinfo {author} {\bibfnamefont {A.}~\bibnamefont
  {Böttcher}}\ and\ \bibinfo {author} {\bibfnamefont {B.}~\bibnamefont
  {Silbermann}},\ }\href {https://doi.org/10.1007/978-1-4612-1426-7} {\emph
  {\bibinfo {title} {Introduction to Large Truncated {Toeplitz} Matrices}}}\
  (\bibinfo  {publisher} {Springer New York},\ \bibinfo {year}
  {1999})\BibitemShut {NoStop}%
\bibitem [{\citenamefont {Nakatsukasa}\ \emph {et~al.}(2018)\citenamefont
  {Nakatsukasa}, \citenamefont {S{\`e}te},\ and\ \citenamefont
  {Trefethen}}]{nakatsukasa2018aaa}%
  \BibitemOpen
  \bibfield  {author} {\bibinfo {author} {\bibfnamefont {Y.}~\bibnamefont
  {Nakatsukasa}}, \bibinfo {author} {\bibfnamefont {O.}~\bibnamefont
  {S{\`e}te}},\ and\ \bibinfo {author} {\bibfnamefont {L.~N.}\ \bibnamefont
  {Trefethen}},\ }\bibfield  {title} {\bibinfo {title} {The {AAA} algorithm for
  rational approximation},\ }\href {https://doi.org/10.1137/16M1106122}
  {\bibfield  {journal} {\bibinfo  {journal} {SIAM J. Sci. Comput.}\ }\textbf
  {\bibinfo {volume} {40}},\ \bibinfo {pages} {A1494} (\bibinfo {year}
  {2018})}\BibitemShut {NoStop}%
\bibitem [{\citenamefont {Sveshnikov}\ and\ \citenamefont
  {Tikhonov}(1978)}]{sveshnikov1978book}%
  \BibitemOpen
  \bibfield  {author} {\bibinfo {author} {\bibfnamefont {A.~G.}\ \bibnamefont
  {Sveshnikov}}\ and\ \bibinfo {author} {\bibfnamefont {A.~N.}\ \bibnamefont
  {Tikhonov}},\ }\href@noop {} {\emph {\bibinfo {title} {The Theory of
  Functions of a Complex Variable}}},\ \bibinfo {edition} {2nd}\ ed.\ (\bibinfo
   {publisher} {Mir Publishers},\ \bibinfo {year} {1978})\BibitemShut {NoStop}%
\bibitem [{\citenamefont {Lee}\ and\ \citenamefont
  {Thomale}(2019)}]{lee2019anatomy}%
  \BibitemOpen
  \bibfield  {author} {\bibinfo {author} {\bibfnamefont {C.~H.}\ \bibnamefont
  {Lee}}\ and\ \bibinfo {author} {\bibfnamefont {R.}~\bibnamefont {Thomale}},\
  }\bibfield  {title} {\bibinfo {title} {Anatomy of skin modes and topology in
  non-{Hermitian} systems},\ }\href
  {https://doi.org/10.1103/PhysRevB.99.201103} {\bibfield  {journal} {\bibinfo
  {journal} {Phys. Rev. B}\ }\textbf {\bibinfo {volume} {99}},\ \bibinfo
  {pages} {201103(R)} (\bibinfo {year} {2019})}\BibitemShut {NoStop}%
\bibitem [{\citenamefont {Zhong}\ \emph {et~al.}(2024)\citenamefont {Zhong},
  \citenamefont {Wang},\ and\ \citenamefont {Fan}}]{zhong2024pole}%
  \BibitemOpen
  \bibfield  {author} {\bibinfo {author} {\bibfnamefont {J.}~\bibnamefont
  {Zhong}}, \bibinfo {author} {\bibfnamefont {H.}~\bibnamefont {Wang}},\ and\
  \bibinfo {author} {\bibfnamefont {S.}~\bibnamefont {Fan}},\ }\bibfield
  {title} {\bibinfo {title} {Pole and zero edge state invariant for
  one-dimensional non-{Hermitian} sublattice symmetry},\ }\href
  {https://doi.org/10.1103/PhysRevB.110.214113} {\bibfield  {journal} {\bibinfo
   {journal} {Phys. Rev. B}\ }\textbf {\bibinfo {volume} {110}},\ \bibinfo
  {pages} {214113} (\bibinfo {year} {2024})}\BibitemShut {NoStop}%
\bibitem [{\citenamefont {Yokomizo}\ and\ \citenamefont
  {Murakami}(2019)}]{yokomizo2019gbz}%
  \BibitemOpen
  \bibfield  {author} {\bibinfo {author} {\bibfnamefont {K.}~\bibnamefont
  {Yokomizo}}\ and\ \bibinfo {author} {\bibfnamefont {S.}~\bibnamefont
  {Murakami}},\ }\bibfield  {title} {\bibinfo {title} {Non-{Bloch} band theory
  of non-{Hermitian} systems},\ }\href
  {https://doi.org/10.1103/PhysRevLett.123.066404} {\bibfield  {journal}
  {\bibinfo  {journal} {Phys. Rev. Lett.}\ }\textbf {\bibinfo {volume} {123}},\
  \bibinfo {pages} {066404} (\bibinfo {year} {2019})}\BibitemShut {NoStop}%
\end{thebibliography}%

\end{document}



\title{
    Detecting non-Hermitian topology from long-time dynamics:\\
    Supplemental Material
}

\date{\today}

\author{Antonio D'Abbruzzo}
\affiliation{The Abdus Salam International Centre for Theoretical Physics, Strada Costiera 11, I-34151 Trieste, Italy}

\author{Rosario Fazio}
\affiliation{The Abdus Salam International Centre for Theoretical Physics, Strada Costiera 11, I-34151 Trieste, Italy}
\affiliation{Dipartimento di Fisica, Università di Napoli ``Federico II'', Monte S. Angelo, I-80126, Napoli, Italy}

\maketitle


The document is structured in two main parts.
In Sec.~\ref{sec:finite_size} we address the questions of how to define the symbol of the covariance matrix for a finite-size system, what is the corresponding evolution equation, and how it differs from the one obtained in the Letter in the thermodynamic limit.
The result is used to lay out how the protocol is practically carried out for finite-size systems.
Instead, in Sec.~\ref{sec:phase_diagram} we illustrate the calculation of the damping matrix for the nonreciprocal Kitaev chain introduced in the Letter, and the exact analytical determination of the associated topological phase diagram.


\section{Finite-size covariance symbol} \label{sec:finite_size}

In order to describe the covariance symbol at finite size, we first review the concept of circulant matrices in Sec.~\ref{subsec:circulant}, which is what we need to formalize the idea of translational invariance under periodic boundary conditions.
Then, in Sec.~\ref{subsec:evolution} we show that the property of being circulant is conserved by the Lindblad dynamics, which allows us to define the finite-size covariance symbol.
We derive the corresponding evolution equation and we show that corrections to the thermodynamic-limit expression are not present at the Fourier points.
This is finally used in Sec.~\ref{subsec:protocol}, where we describe how our protocol is practically carried out through Fourier-point probing, rational extrapolation, and robust winding number calculation.


\subsection{Circulant matrices} \label{subsec:circulant}

Let $N$ and $b$ be two positive integers.
We say that $A \in \mathbb{C}^{Nb \times Nb}$ is a \textit{(block-)circulant matrix} when it can be written in the form~\cite{bottcher1999book}
\begin{equation} \label{eq:kappa_circulant}
	A = \sum_{n \in \mathbb{T}_N} S^n \otimes A^{(n)},
	\qquad
	S := \begin{pmatrix}
		0 & 1 & 0 & \ldots & 0 \\
		0 & 0 & 1 & \ldots & 0 \\
		\vdots & \vdots & \vdots & \ddots & \vdots \\
		0 & 0 & 0 & \ldots & 1 \\
		1 & 0 & 0 & \ldots & 0
	\end{pmatrix} \in \mathbb{R}^{N \times N},
	\qquad
	A^{(n)} \in \mathbb{C}^{b \times b},
\end{equation}
where $\mathbb{T}_N := \qty{-\lfloor\frac{N-1}{2}\rfloor, \ldots, \lceil\frac{N-1}{2}\rceil}$ is a range of $N$ consecutive integers centered around zero.
The matrix $S$ is called \textit{cyclic shift}, and it satisfies $S^N = \mathbbm{1}_N$.
The following is an alternative characterization.

\begin{lemma} \label{lemma:commutator}
	$A \in \mathbb{C}^{Nb \times Nb}$ is circulant if and only if $[A, S \otimes \mathbbm{1}_b] = 0$.
\end{lemma}
\begin{proof}
    For each $i, j \in \qty{1, \ldots, N}$, let $E^{(i,j)}$ be the $N \times N$ matrix having 1 in position $(i,j)$ and 0 elsewhere.
    The cyclic shift can then be written in the form $S = \sum_{k=1}^N E^{(k,k+1)}$, where it is understood that indices wrap around when they go beyond the borders of $\qty{1, \ldots, N}$.
    Since $E^{(i,j)} E^{(k,l)} = \delta_{jk} E^{(i,l)}$, one easily see by induction that $S^n = \sum_{k=1}^N E^{(k,k+n)}$.
    Also, any matrix $A \in \mathbb{C}^{Nb \times Nb}$ can be uniquely written in the block form $A = \sum_{i,j=1}^N E^{(i,j)} \otimes A^{(i,j)}$ where $A^{(i,j)} \in \mathbb{C}^{b \times b}$.

    Suppose $A$ is circulant, so that $A = \sum_{n \in \mathbb{T}_N} S^n \otimes A^{(n)}$.
    For each $n$, we obviously have $[S^n \otimes A^{(n)}, S \otimes \mathbbm{1}_b] = 0$, hence $[A, S \otimes \mathbbm{1}_b] = 0$.
    On the other hand, suppose $[A, S \otimes \mathbbm{1}_b] = 0$ for a general block matrix $A$.
    We can write $A(S \otimes \mathbbm{1}_b) = \sum_{i,j=1}^N E^{(i,j)} S \otimes A^{(i,j)} = \sum_{i,j=1}^N E^{(i,j+1)} \otimes A^{(i,j)} = \sum_{i,j=1}^N E^{(i,j)} \otimes A^{(i,j-1)}$.
    Similarly, for the other product, we can write $(S \otimes \mathbbm{1}_b)A = \sum_{i,j=1}^N E^{(i,j)} \otimes A^{(i+1,j)}$.
    By the uniqueness of the block decomposition, we must have $A^{(i,j-1)} = A^{(i+1,j)}$, which is equivalent to say that $A^{(i,j)}$ actually depends only on $i-j$ modulo $N$.
    If we define $A^{(n)} := A^{(1,1+n)}$ we now have $\sum_{n \in \mathbb{T}_N} S^n \otimes A^{(n)} = \sum_{n \in \mathbb{T}_N} \sum_{k=1}^N E^{(k,k+n)} \otimes A^{(k,k+n)} = A$, by uniqueness of the block decomposition and the fact that the pair $(k,k+n)$ runs over $\qty{1, \ldots, N}^2$ exactly once.
\end{proof}

Let $A$ be a circulant matrix in the form~\eqref{eq:kappa_circulant}.
Its \textit{symbol} is the function $\mathcal{A}: \mathbb{C} \setminus \qty{0} \to \mathbb{C}^{b \times b}$ defined by~\cite{bottcher1999book}
\begin{equation}
	\mathcal{A}(z) := \sum_{n \in \mathbb{T}_N} A^{(n)} z^n.
\end{equation}
This is a Laurent polynomial in $z$ with matrix coefficients.
We will need the following algebraic property of symbols.

\begin{lemma} \label{lemma:product_symbols}
	Let $A = \sum_{n \in \mathbb{T}_N} S^n \otimes A^{(n)}$ and $B = \sum_{n \in \mathbb{T}_N} S^n \otimes B^{(n)}$ be two circulant matrices, and let $C = AB$.
	The symbols of these matrices are related by
	\begin{equation}
		\mathcal{C}(z) \equiv \mathcal{A}(z) \mathcal{B}(z) \mod \qty(z^N - 1).
	\end{equation}
\end{lemma}
\begin{proof}
	Given $n,m \in \mathbb{T}_N$ we can always fold back the sum $n+m$ into $\mathbb{T}_N$ by writing $n+m = \expval{n+m} + N q(n,m)$, where $\expval{n+m} \in \mathbb{T}_N$ and $q(n,m) := (n+m - \expval{n+m})/N \in \qty{-1, 0, 1}$.
	Therefore, since $S^N = \mathbbm{1}_N$,
	\begin{equation}
		C = \sum_{n,m \in \mathbb{T}_N} S^{n+m} \otimes A^{(n)} B^{(m)}
		= \sum_{k \in \mathbb{T}_N} S^k \otimes C^{(k)},
		\qquad
		C^{(k)} := \sum_{\substack{n,m \in \mathbb{T}_N \\ \expval{n+m} = k}} A^{(n)} B^{(m)}.
	\end{equation}
	Clearly, $\mathcal{C}(z) = \sum_{k \in \mathbb{T}_N} C^{(k)} z^k$.
	On the other hand,
	\begin{equation}
		\mathcal{A}(z) \mathcal{B}(z) = \sum_{n,m \in \mathbb{T}_N} A^{(n)} B^{(m)} z^{n+m}
		= \sum_{k \in \mathbb{T}_N} \sum_{\substack{n,m \in \mathbb{T}_N \\ \expval{n+m} = k}} A^{(n)} B^{(m)} z^{N q(n,m)} z^k.
	\end{equation}
	By writing $z^{N q(n,m)} = \qty( z^{N q(n,m)} - 1 ) + 1$, we see that $\mathcal{C}(z) = \mathcal{A}(z) \mathcal{B}(z) + \Delta(z)$ with
	\begin{equation}
		\Delta(z) := \sum_{n,m \in \mathbb{T}_N} A^{(n)} B^{(m)} \qty(1 - z^{N q(n,m)}) z^{\expval{n+m}}.
	\end{equation}
	This is what we wanted to prove, since $\Delta(z) = 0$ when $z^N = 1$.
\end{proof}


\subsection{Covariance symbol evolution} \label{subsec:evolution}

Consider a finite-size translationally-invariant system under periodic boundary conditions.
In the context of open Gaussian systems, this is equivalent to say that the matrices $X$ and $Y$ appearing in the evolution equation of the covariance matrix, $\dot{\Gamma} = X \Gamma + \Gamma X^T + Y$, are circulant in the sense defined above.
The following theorem establishes that in this case the covariance inherits the translational invariance.
\begin{theorem}
	If $X$, $Y$, and $\Gamma(0)$ are circulant matrices, then the covariance $\Gamma(t)$ and the deviation from the steady state $G(t) := \Gamma(t) - \Gamma_\infty$ are circulant matrices for all times $t \geq 0$.
\end{theorem}
\begin{proof}
    According to Lemma~\ref{lemma:commutator}, we can prove that $\Gamma(t)$ is circulant by showing that $F(t) := [\Gamma(t), \mathbb{S}] = 0$ for all times $t \geq 0$, where $\mathbb{S} := S \otimes \mathbbm{1}_2$ and $S$ is the cyclic shift.
    First, note that if $X$ commutes with $\mathbb{S}$ it also commutes with any power of $\mathbb{S}$; in particular, it commutes with $\mathbb{S}^{N-1} = \mathbb{S}^T$, which is equivalent to say that $[X^T, \mathbb{S}] = 0$.
    Now, using the facts that $F(0) = 0$ and $[X, \mathbb{S}] = [X^T, \mathbb{S}] = [Y, \mathbb{S}] = 0$,
    \begin{align}
    	\dot{F} &= \dot{\Gamma} \mathbb{S} - \mathbb{S} \dot{\Gamma}
    	= (X \Gamma + \Gamma X^T + Y) \mathbb{S} - \mathbb{S} (X \Gamma + \Gamma X^T + Y) \nonumber \\
    	&= X \Gamma \mathbb{S} - \mathbb{S} X \Gamma + \Gamma X^T \mathbb{S} - \mathbb{S} \Gamma X^T
    	= X \Gamma \mathbb{S} - X \mathbb{S} \Gamma + \Gamma \mathbb{S} X^T - \mathbb{S} \Gamma X^T \\
    	&= X F + F X^T \qquad \Rightarrow \qquad F(t) = e^{X t} F(0) e^{X^T t} = 0. \nonumber
    \end{align}
    This proves that $\Gamma(t)$ is circulant.
    For what concerns the steady state, start from $X \Gamma_\infty + \Gamma_\infty X^T + Y = 0$.
    Multiplying on the left by $\mathbb{S}$ we have $0 = \mathbb{S} X \Gamma_\infty + \mathbb{S} \Gamma_\infty X^T + \mathbb{S} Y = X \mathbb{S} \Gamma_\infty + \mathbb{S} \Gamma_\infty X^T + \mathbb{S} Y$.
    Multiplying on the right by $\mathbb{S}$, we obtain instead $0 = X \Gamma_\infty \mathbb{S} + \Gamma_\infty X^T \mathbb{S} + Y \mathbb{S} = X \Gamma_\infty \mathbb{S} + \Gamma_\infty \mathbb{S} X^T + Y \mathbb{S}$.
    Subtracting the two equations, we end up with $X [\mathbb{S}, \Gamma_\infty] + [\mathbb{S}, \Gamma_\infty] X^T = 0$, which gives $[\mathbb{S}, \Gamma_\infty] = 0$ under the same hypotheses guaranteeing the existence of $\Gamma_\infty$.
    We conclude that $\Gamma_\infty$ is circulant, and therefore the deviation $G(t)$ is circulant too at all times $t \geq 0$.
\end{proof}

Thanks to this result, if our (finite-size) system has periodic boundary conditions and starts out in a translationally invariant state, then it makes sense to consider the symbol of the covariance deviation $\mathcal{G}(z, t) := \sum_{n \in \mathbb{T}_N} G^{(n)}(t) z^n$.
Since $\dot{G} = X G + G X^T$, Lemma~\ref{lemma:product_symbols} also gives us immediately the evolution equation
\begin{equation} \label{eq:symbol_evolution}
	\dv{t} \mathcal{G}(z, t) \equiv \mathcal{X}(z) \mathcal{G}(z, t) + \mathcal{G}(z, t) \mathcal{X}^T(z^{-1}) \mod \qty(z^N - 1).
\end{equation}
In case $z^N = 1$, this is the same equation obtained in the Letter for the infinite system.
Otherwise, finite-size corrections appear according to what established by Lemma~\ref{lemma:product_symbols}.
Note that imposing $z^N = 1$ is equivalent to require $z = z_k := e^{2 \pi \im k/N}$ with $k \in \qty{1, \ldots, N}$.
These are the usual Fourier points, and taking the symbol when $z = z_k$ is equivalent to take the discrete Fourier transform.


\subsection{Description of the protocol} \label{subsec:protocol}

We now briefly describe how the protocol to detect non-Hermitian topology is carried out starting from covariance deviation data $G(t)$ associated with a finite-size system under periodic boundary conditions.

According to Eq.~\eqref{eq:symbol_evolution}, the discrete Fourier transform $\mathcal{G}(z_k, t)$ of $G(t)$ follows the same evolution as the infinite-size symbol.
Therefore, as described in the Letter, the first step is to calculate $\mathcal{G}(z_k, t)$ for $k = 1, \ldots, N$ and the ratios $f(z_k, t) := [\mathcal{G}_{11}(z_k, t) / \mathcal{G}_{21}(z_k, t)]^2$ at large $t$.
Except for possibly few of the points $z_k$, where no dominant contribution to the ratio may emerge, $f(z_k, t)$ should reproduce the analytical prediction $f(z_k) := \alpha(z_k) / \beta(z_k)$.

In order to calculate the topological invariant, we need $f(z) := \alpha(z)/\beta(z)$ for $z \neq z_k$.
Since this is a ratio between two finite-order polynomials, from the numerical point of view the problem can be tackled efficiently using rational approximation algorithms, such as the standard AAA algorithm~\cite{nakatsukasa2018aaa}.
Basically, it takes as input some ``support points'' where a function is known (in our case, the values at $z_k$) and extrapolates it to other points using a ratio-of-polynomials representation.
In order to achieve reasonable accuracy, a sufficiently large number of support points is required.
In our case, $\alpha(z)$ and $\beta(z)$ are known to be Laurent polynomials with exponents in a certain range $[-R, R]$.
Therefore, by setting up the system $\alpha(z_k) = f(z_k) \beta(z_k)$, it is theoretically possible to infer the coefficients of $\alpha(z)$ and $\beta(z)$ exactly, and hence find $f(z)$, with $N \geq 4R+1$ support points.
In practice, a larger number is required because the values at the support points do not coincide exactly with $f(z_k)$, for a variety of reasons: the covariance deviation $G(t)$ may be affected by noise; the time $t$ is finite and convergence to $f(z_k)$ is only asymptotic; at some $z_k$ the data may not converge at all.
Also, upon increasing $N$, one should be mindful about numerical artifacts known as Froissart doublets, a common source of issues for rational approximation algorithms~\cite{nakatsukasa2018aaa}.
Basically, a Froissart doublet is a root and a pole of a computed rational approximant that lie very close to each other and that do not belong to the true function.
Their presence produces local damages to the approximation, showing up as ``outliers'' with respect to the true function when extrapolating near that points.
While modern algorithms are designed to detect and clean up as much Froissart doublets as possible, their apparance cannot be ruled out.

Once $f(z)$ has been determined, we need to calculate its winding $\nu(r)$ around circles of radius $r$, and then we should read its value after $2R$ jumps occurred upon sweeping $r$.
The straightfoward method to calculate $\nu(r)$ is a simple phase unwrapping, which follows from the formula
\begin{equation}
	\nu(r) := \frac{1}{4 \pi \im} \oint_{|z| = r} \dd{\log f(z)}
	= \frac{1}{4 \pi \im} \int_0^{2 \pi} \partial_\theta \log f(r e^{\im \theta}) \dd{\theta}
	= \frac{1}{4 \pi} \int_0^{2 \pi} \partial_\theta \arg f(r e^{\im \theta}) \dd{\theta},
\end{equation}
where we used $\dd{z} = \im r e^{\im \theta} \dd{\theta}$ when $z = r e^{\im \theta}$, and the fact that $\log|f(r e^{\im \theta})|$ is single-valued and $2\pi$-periodic in $\theta$.
However, outliers caused by Froissart doublets or non-converging points may severely disrupt this calculation and may lead to relevant quantized errors.
The problem can be circumvented by recalling the polar Cauchy-Riemann equation~\cite{sveshnikov1978book} applied to $\log f$:
\begin{equation}
	\partial_\theta \arg f = r \partial_r \log |f|.
\end{equation}
If we define the mean log-magnitude
\begin{equation}
	g(r) := \frac{1}{2\pi} \int_0^{2\pi} \log |f(r e^{\im \theta})| \dd{\theta},
\end{equation}
we can then see that $\nu(r)$ is determined by the logarithmic derivative of $g(r)$:
\begin{equation}
	\frac{1}{2} \dv{g(r)}{\log r} = \frac{r}{2} \dv{g(r)}{r} = \frac{r}{4 \pi} \int_0^{2\pi} \partial_r \log |f(r e^{\im \theta})| \dd{\theta}
	= \frac{1}{4 \pi} \int_0^{2\pi} \partial_\theta \arg f(r e^{\im \theta}) \dd{\theta} = \nu(r).
\end{equation}
The advantage of using this formula is that $g(r)$ is robust against fluctuations in $|f|$, and outliers in the data show up as noise in $\nu(r)$ instead of discrete jumps.


\section{Phase diagram of the nonreciprocal Kitaev chain} \label{sec:phase_diagram}

In this section we analytically determine the point-gap topological phase diagram of the nonreciprocal Kitaev chain introduced in the Letter.
First, we recall how the model is defined, we move to the Majorana representation, and we derive the corresponding damping matrix.
Then, we identify the functions $\alpha(z)$ and $\beta(z)$ and we carry out the computation of the topological invariant $\nu$ exploiting their polynomial structure.

We start from the following Hamiltonian $\hat H$ and Lindblad operators $\hat L_j$, which are here written using the fermionic ladder operators $\qty{\hat a_j}_{j=1}^N$:
\begin{subequations}
\begin{align}
	\hat H &= \mu \sum_j \hat a_j^\dagger \hat a_j + \sum_j \qty[
	    J \hat a_j^\dagger \hat a_{j+1} + \Delta \hat a_j^\dagger \hat a_{j+1}^\dagger + \text{H.c.}
	], \\
	\hat L_j &= \sqrt{\gamma} \qty(\hat a_j + \im \, \hat a_{j+1}).
\end{align}
\end{subequations}
We assume open boundary conditions, so that $\hat a_0 \equiv \hat a_{N+1} \equiv 0$.
Also, we assume the coefficients $\mu, J, \Delta, \gamma$ to be real parameters.
In the closed-system case, $\gamma = 0$, it is known that the chain has nontrivial topology for $|\mu| < 2|J|$, assuming $\Delta \neq 0$.
For simplicity, it is therefore natural to fix $J, \Delta > 0$ and to study the phase diagram in the plane $(\gamma, \mu)$.
Also, $\gamma \geq 0$ holds, since it is defined as a dissipation rate.

In terms of the Majorana operators $\hat w_j^A := \frac{1}{\sqrt 2} (\hat a_j^\dagger + \hat a_j)$, $\hat w_j^B := \frac{\im}{\sqrt 2} (\hat a_j^\dagger - \hat a_j)$, we find
\begin{subequations}
\begin{align}
	\hat H &= \frac{\im}{2} \sum_j \qty[ \mu \, \hat w_j^A \hat w_j^B + (J-\Delta) \hat w_j^A \hat w_{j+1}^B + (J+\Delta) \hat w_{j+1}^A \hat w_j^B - \text{H.c.} ], \\
	\hat L_j &= \sqrt{\frac{\gamma}{2}} \qty( \hat w_j^A + \im \, \hat w_j^B + \im \, \hat w_{j+1}^A - \hat w_{j+1}^B ).
\end{align}
\end{subequations}
Using the notation introduced in the Letter, we group the Majorana operators in the $2N$-dimensional vector ${\hat{\mathbf w}^\dagger = (\hat w_1^A, \dots, \hat w_N^A, \hat w_1^B, \dots, \hat w_N^B)}$.
In terms of it, we write $\hat H = \frac{\im}{2} \hat{\mathbf w}^\dagger H \hat{\mathbf w}$ and $\hat{\mathbf L} = L \hat{\mathbf w}$ with
\begin{subequations}
\begin{equation}
    H = \left(\begin{array}{ccccc|ccccc}
      0 & 0 & \cdots & 0 & 0 & \mu & J-\Delta & 0 & \cdots & 0 \\
      0 & 0 & \ddots & \vdots & \vdots & J+\Delta & \mu & J-\Delta & \ddots & \vdots \\
      \vdots & \ddots & \ddots & 0 & 0 & 0 & J+\Delta & \mu & \ddots & 0 \\
      0 & \cdots & 0 & 0 & 0 & \vdots & \ddots & \ddots & \ddots & J-\Delta \\
      0 & \cdots & 0 & 0 & 0 & 0 & \cdots & 0 & J+\Delta & \mu \\
      \hline
      -\mu & -(J+\Delta) & 0 & \cdots & 0 & 0 & 0 & \cdots & 0 & 0 \\
      -(J-\Delta) & -\mu & -(J+\Delta) & \ddots & \vdots & 0 & 0 & \ddots & \vdots & \vdots \\
      0 & -(J-\Delta) & -\mu & \ddots & 0 & \vdots & \ddots & \ddots & 0 & 0 \\
      \vdots & \ddots & \ddots & \ddots & -(J+\Delta) & 0 & \cdots & 0 & 0 & 0 \\
      0 & \cdots & 0 & -(J-\Delta) & -\mu & 0 & \cdots & 0 & 0 & 0
    \end{array}\right) \in \mathbb{R}^{2N \times 2N}, \\
\end{equation}
\begin{equation}
    L = \sqrt{\frac{\gamma}{2}}
    \left(\begin{array}{ccccc|ccccc}
    \im & 0 & 0 & \cdots & 0 & -1 & 0 & 0 & \cdots & 0 \\
    1 & \im & 0 & \ddots & \vdots & \im & -1 & 0 & \ddots & \vdots \\
    0 & 1 & \im & \ddots & 0 & 0 & \im & -1 & \ddots & 0 \\
    \vdots & \ddots & \ddots & \ddots & \vdots & \vdots & \ddots & \ddots & \ddots & \vdots \\
    0 & \cdots & 0 & 1 & \im & 0 & \cdots & 0 & \im & -1 \\
    0 & \cdots & 0 & 0 & 1 & 0 & \cdots & 0 & 0 & \im
    \end{array}\right) \in \mathbb{C}^{(N+1) \times 2N}.
\end{equation}
\end{subequations}
We then arrive at the damping matrix $X := H - \Re(L^\dagger L)$:
\begin{equation}
	X = \left(\begin{array}{ccccc|ccccc}
      -\gamma & 0 & \cdots & 0 & 0 & \mu & J_+ - \Delta & 0 & \cdots & 0 \\
      0 & -\gamma & \ddots & \vdots & \vdots & J_- + \Delta & \mu & J_+ - \Delta & \ddots & \vdots \\
      \vdots & \ddots & \ddots & 0 & 0 & 0 & J_- + \Delta & \mu & \ddots & 0 \\
      0 & \cdots & 0 & -\gamma & 0 & \vdots & \ddots & \ddots & \ddots & J_+ - \Delta \\
      0 & \cdots & 0 & 0 & -\gamma & 0 & \cdots & 0 & J_- + \Delta & \mu \\
      \hline
      -\mu & -(J_+ + \Delta) & 0 & \cdots & 0 & -\gamma & 0 & \cdots & 0 & 0 \\
      -(J_- - \Delta) & -\mu & -(J_+ + \Delta) & \ddots & \vdots & 0 & -\gamma & \ddots & \vdots & \vdots \\
      0 & -(J_- - \Delta) & -\mu & \ddots & 0 & \vdots & \ddots & \ddots & 0 & 0 \\
      \vdots & \ddots & \ddots & \ddots & -(J_+ + \Delta) & 0 & \cdots & 0 & -\gamma & 0 \\
      0 & \cdots & 0 & -(J_- - \Delta) & -\mu & 0 & \cdots & 0 & 0 & -\gamma
    \end{array}\right),
\end{equation}
where $J_\pm := J \pm \gamma/2$.
We also read off the symbol of $X$ as
\begin{equation}
	\mathcal{X}(z) = \begin{pmatrix}
		-\gamma & \alpha(z) \\
		\beta(z) & -\gamma
	\end{pmatrix},
	\qquad
	\begin{cases}
		\alpha(z) = (J_- + \Delta) z^{-1} + \mu + (J_+ - \Delta) z, \\
		\beta(z) = -(J_- - \Delta) z^{-1} - \mu - (J_+ + \Delta) z.
	\end{cases}
\end{equation}

As we said in the Letter, the topological invariant $\nu = \frac{1}{4 \pi \im} \oint \dd{\log f(z)}$ can be calculated from the polynomial structure of $f(z) := \alpha(z)/\beta(z)$.
If $\qty{z_i}_{i=1}^4$ is the set of zeros and poles of $f(z)$ ordered by magnitude, it can be proved that $\nu$ is the half difference between the number of zeros and poles of $f(z)$ contained in the set $\qty{z_1, z_2}$~\cite{lee2019anatomy, zhong2024pole}.
In our case, the zeros $z^\alpha_\pm$ of $f(z)$ are the roots of $(J_+ - \Delta) z^2 + \mu z + (J_- + \Delta) = 0$, while the poles $z^\beta_\pm$ of $f(z)$ are the roots of $(J_+ + \Delta) z^2 + \mu z + (J_- - \Delta) = 0$.
Explicitly,
\begin{equation}
	z^\alpha_\pm = \frac{-\mu \pm \sqrt{\Omega_\alpha}}{2(J_+ - \Delta)},
	\qquad
	z^\beta_\pm = \frac{-\mu \pm \sqrt{\Omega_\beta}}{2(J_+ + \Delta)},
\end{equation}
where we defined $\Omega_\alpha := \mu^2 - 4(J_+ - \Delta)(J_- + \Delta)$ and $\Omega_\beta := \mu^2 - 4(J_+ + \Delta)(J_- - \Delta)$.
Equivalently,
\begin{equation}
	\Omega_\alpha = \mu^2 + (\gamma - 2 \Delta)^2 - 4 J^2,
	\qquad
	\Omega_\beta = \mu^2 + (\gamma + 2 \Delta)^2 - 4 J^2.
\end{equation}
In the $(\gamma, \mu)$ plane, the conditions $\Omega_\alpha = 0$ and $\Omega_\beta = 0$ respectively identify circles of radius $2J$ centered in $(\pm 2 \Delta, 0)$.
Since $\Delta > 0$, the portion of the $\Omega_\beta$ circle with $\gamma \geq 0$ is entirely contained in the portion of the $\Omega_\alpha$ circle with $\gamma \geq 0$, and is relevant only when $\Delta < J$.
It will also be useful to recall the following Vieta's formulas:
\begin{equation}
	|z^\alpha_- z^\alpha_+| = \kappa_\alpha^2,
	\quad
	\kappa_\alpha := \sqrt{\abs{\frac{J_- + \Delta}{J_+ - \Delta}}},
	\qquad
	|z^\beta_- z^\beta_+| = \kappa_\beta^2,
	\quad
	\kappa_\beta := \sqrt{\abs{\frac{J_- - \Delta}{J_+ + \Delta}}}.
\end{equation}
Note that when $\mu = 0$ or $\Omega_\alpha \leq 0$ the two $\alpha$ roots are conjugate and share the same magnitude $|z^\alpha_\pm| = \kappa_\alpha$, given by the Vieta factor.
Instead, when $\mu \neq 0$ and $\Omega_\alpha > 0$ the two $\alpha$ roots are real and have different magnitude.
Specifically, when $\mu > 0$ it holds $|z^\alpha_+| < |z^\alpha_-|$, while the opposite is true for $\mu < 0$.
Similar statements hold for the $\beta$ roots as well.

Our objective now is to order these roots by magnitude and to identify how many of the first two are $\alpha$ roots or $\beta$ roots: this counting immediately gives the winding number we are looking for.
To do that, we subdivide the $(\gamma, \mu)$ plane into the following nonoverlapping regions:
\begin{equation}
	\mathcal{P}_1 := \qty{(\gamma, \mu) : \Omega_\beta \leq 0}, \qquad
	\mathcal{P}_2 := \qty{(\gamma, \mu) : \Omega_\beta > 0, \, \Omega_\alpha \leq 0}, \qquad
	\mathcal{P}_3 := \qty{(\gamma, \mu) : \Omega_\alpha > 0}.
\end{equation}
In each of them we have a definite internal ordering between the two $\alpha$ roots and the two $\beta$ roots, and we only need to understand the relative ordering between the two categories.

First, we handle once and for all the special case $\mu = 0$.
Here the roots have magnitudes given by their respective Vieta factors.
A straightforward computation reveals that $\kappa_\alpha < \kappa_\beta$ is equivalent to $\gamma > 2 \sqrt{J^2 + \Delta^2}$, therefore
\begin{subequations}
\begin{align}
	\mu = 0, \, 0 \leq \gamma < 2 \sqrt{J^2 + \Delta^2}
	\quad&\Rightarrow\quad \nu = -1, \\
	\mu = 0, \, \gamma > 2 \sqrt{J^2 + \Delta^2}
	\quad&\Rightarrow\quad \nu = 1.
\end{align}
\end{subequations}
From now on, we assume $\mu \neq 0$.

It is easy to see what happens in $\mathcal{P}_1$ (which is nonempty only when $\Delta < J$).
The roots again have magnitudes given by the Vieta factors, but since we are inside the $\Omega_\beta$ circle we must have $\gamma \leq 2(J-\Delta) < 2 \sqrt{J^2 + \Delta^2}$, therefore $\kappa_\beta < \kappa_\alpha$ and $\nu = -1$ in the entirety of $\mathcal{P}_1$.

Now we focus on $\mathcal{P}_2$.
The $\alpha$ roots still have magnitudes equal to $\kappa_\alpha$, while the $\beta$ roots are splitted.
There are three possible cases according to where $\kappa_\alpha$ sits: before, after, or between the $\beta$ roots.
We study these cases separately.
Since the ordering between $z^\beta_+$ and $z^\beta_-$ is dictated by the sign of $\mu$, we use the compact notations $z^\beta_\pm$ and $z^\beta_\mp$: it is understood that the upper sign is for $\mu > 0$ while the lower sign is for $\mu < 0$.
Also, we will make use of some technical lemmas whose proof is postponed to Sec.~\ref{subsec:technicals}.
\begin{itemize}
	\item $\kappa_\alpha < |z^\beta_\pm|$, so that $\nu = 1$.
	The inequality has the form
	\begin{equation} \label{eq:temp1}
		a < \abs{-\mu \pm \sqrt{\mu^2 + b}}
		= \abs{\sqrt{\mu^2 + b} - \abs{\mu}}
		= \mathrm{sgn}(b) \qty[\sqrt{\mu^2 + b} - |\mu|],
	\end{equation}
	where $a := 2 \kappa_\alpha |J_+ + \Delta| > 0$ and $b := (\gamma + 2 \Delta)^2 - 4J^2$.
	In Lemma~\ref{lemma:ineq1} below, we prove that a solution for $\mu$ exists only when $a^2 < |b|$ and is $|\mu| < \mathrm{sgn}(b) (b-a^2)/2a$.
	A computation reveals that in our case $a^2 < |b|$ is equivalent to $\gamma > 2 \sqrt{J^2 + \Delta^2}$, and that the solution writes
	\begin{equation}
		|\mu| < \frac{\gamma^2 - 4(J^2 + \Delta^2)}{\sqrt{4J^2 - (\gamma - 2\Delta)^2}}.
	\end{equation}
	It is also important to note that the boundary of the above region intersects the $\Omega_\alpha$ circle at $\gamma = \Delta + \sqrt{\Delta^2 + 4J^2}$.
	\item $|z^\beta_\mp| < \kappa_\alpha$, so that $\nu = -1$.
	The inequality has the form
	\begin{equation}
	    a > \abs{-\mu \mp \sqrt{\mu^2 + b}} = \sqrt{\mu^2 + b} + |\mu|,
	\end{equation}
	where $a$ and $b$ are the same as for Eq.~\eqref{eq:temp1}.
	In Lemma~\ref{lemma:ineq2} below, we prove that this time a solution for $\mu$ exists only when $|b| < a^2$, and is $|\mu| < (a^2 - b)/2a$.
	We therefore found the region with $\gamma < 2 \sqrt{J^2 + \Delta^2}$ and
	\begin{equation}
		|\mu| < \frac{4(J^2 + \Delta^2) - \gamma^2}{\sqrt{4J^2 - (\gamma - 2\Delta)^2}}.
	\end{equation}
	Note also the boundary of the above region intersects the $\Omega_\alpha$ circle at $\gamma = 2 \Delta$.
	\item $|z^\beta_\pm| < \kappa_\alpha < |z^\beta_\mp|$.
	In this case, $z^\alpha_\pm$ belong to the GBZ, hence $\alpha(z)$ becomes singular on the contour and we hit an exceptional point of the damping matrix~\cite{yokomizo2019gbz}.
	In this case, the winding number $\nu$ is not well defined.
	This covers the complementary region of the ones we found in the previous points of the list.
\end{itemize}
With the three cases above we treated the entirety of $\mathcal{P}_2$.

We conclude the study of the phase diagram by focusing on $\mathcal{P}_3$.
Assuming $\mu \neq 0$, the $\alpha$ roots and the $\beta$ roots are all splitted now, and there are again three cases to consider.
\begin{itemize}
	\item The $\alpha$ roots stay below the $\beta$ roots, that is $|z^\alpha_\mp| < |z^\beta_\pm|$ and $\nu = 1$.
	The inequality is of the form
	\begin{equation}
		\abs{-\mu \pm \sqrt{\mu^2 + b}} > c \abs{-\mu \mp \sqrt{\mu^2 + d}}
		\quad\Rightarrow\quad
		\mathrm{sgn}(b) \qty[\sqrt{\mu^2 + b} - |\mu|] > c \qty[\sqrt{\mu^2 + d} + |\mu|],
	\end{equation}
	where $b = (\gamma + 2\Delta)^2 - 4J^2$ as before and we additionally defined $d := (\gamma - 2\Delta)^2 - 4J^2$ and $c := |J_+ + \Delta|/|J_+ - \Delta|$.
	The solution is reported in Lemma~\ref{lemma:ineq3} below.
	Plugging our definitions for the parameters, one finds
	\begin{equation}
		\gamma > \Delta + \sqrt{\Delta^2 + 4J^2}, \qquad
		|\mu| < \frac{4 J \Delta}{\sqrt{\gamma^2 - 4J^2}},
	\end{equation}
	which should also be intersected with $\Omega_\alpha > 0$.
	\item The $\beta$ roots stay below the $\alpha$ roots, that is $|z^\beta_\mp| < |z^\alpha_\pm|$ and $\nu = -1$.
	Now the inequality reads
	\begin{equation}
		\abs{-\mu \mp \sqrt{\mu^2 + b}} < c \abs{-\mu \pm \sqrt{\mu^2 + d}}
		\quad\Rightarrow\quad
		\sqrt{\mu^2 + b} + |\mu| < c \, \mathrm{sgn}(d) \qty[\sqrt{\mu^2 + d} - |\mu|].
	\end{equation}
	The solution to is reported in Lemma~\ref{lemma:ineq4} below.
	With our parameters one arrives at the region
	\begin{equation}
		\gamma < 2\Delta, \qquad |\mu| < 2J.
	\end{equation}
	\item The $\alpha$ and $\beta$ roots are alternated, and $\nu = 0$.
	This covers all the region in $\mathcal{P}_3$ that is not already included by the results obtained in the previous points of the list.
\end{itemize}

Putting together the various pieces of the computation, we arrive at the phase diagram illustrated in the Letter.


\subsection{Some technical lemmas} \label{subsec:technicals}

\begin{lemma} \label{lemma:ineq1}
	Let $a, b \in \mathbb{R}$ such that $a > 0$ and $b \neq 0$.
	Assuming $x^2 + b \geq 0$, the inequality
	\begin{equation}
		a < \mathrm{sgn}(b) \qty[\sqrt{x^2 + b} - |x|]
	\end{equation}
	admits solutions only when $a^2 < |b|$, and in this case the solution set is
	\begin{equation}
		|x| < \mathrm{sgn}(b) \frac{b-a^2}{2a}.
	\end{equation}
\end{lemma}
\begin{proof}
	Suppose first $b > 0$, so that $\sqrt{x^2 + b} > a + |x|$.
	Both sides are positive, and after squaring them we find $|x| < (b-a^2)/2a$.
	Since $|x|$ is positive, this is valid only when $a^2 < b$.

	Suppose instead $b < 0$, so that the inequality is $\sqrt{x^2 + b} < |x| - a$.
	Note that this requires $|x| > a$.
	Squaring both sides yields $|x| < (a^2 - b)/2a$.
	Now, note that the function $|x| - \sqrt{x^2 + b}$ is bounded above by $\sqrt{-b}$, a value attained for $|x| = \sqrt{-b}$.
	So, we must require $-b > a^2$.
	The condition $|x| > a$ is then automatic.
\end{proof}

\begin{lemma} \label{lemma:ineq2}
	Let $a, b \in \mathbb{R}$ such that $a > 0$ and $b \neq 0$.
	Assuming $x^2 + b \geq 0$, the inequality
	\begin{equation}
		a > \sqrt{x^2 + b} + |x|
	\end{equation}
	admits solutions only when $|b| < a^2$, and in this cae the solution set is
	\begin{equation}
		|x| < \frac{a^2 - b}{2a}.
	\end{equation}
\end{lemma}
\begin{proof}
	The inequality can be rewritten as $a-|x| > \sqrt{x^2 + b}$, so we need to require $|x| < a$.
	Squaring both sides we obtain $|x| < (a^2 - b)/2a$, which requires $b < a^2$.
	In case $b > 0$, $(a^2-b)/2a < a$ so that $|x| < a$ is automatic.
	In case $b < 0$, we must additionally require $|x| \geq \sqrt{-b}$, and this is compatible with $|x| < a$ only when $-b < a^2$.
\end{proof}

\begin{lemma} \label{lemma:ineq3}
	Let $b, c, d \in \mathbb{R}$ such that $c > 0$ and $b, d \neq 0$.
	Assuming $x^2 + b \geq 0$ and $x^2 + d \geq 0$, the inequality
	\begin{equation} \label{eq:ineq3}
		\mathrm{sgn}(b) \qty[\sqrt{x^2 + b} - |x|] > c \qty[\sqrt{x^2 + d} + |x|]
	\end{equation}
	admits solutions only when
	\begin{equation}
		1 + \mathrm{sgn}(b)c > 0, \qquad
		|d| < \frac{cd + |b|}{c(1 + \mathrm{sgn}(b)c)} < \frac{|b|}{c^2},
	\end{equation}
	and in this case the solution set is
	\begin{equation}
		|x| < \frac{|b-c^2 d|}{2 \sqrt{c(1 + \mathrm{sgn}(b)c)(cd + |b|)}}.
	\end{equation}
\end{lemma}
\begin{proof}
    The left-hand side of the inequality is equal to $|b|/[\sqrt{x^2 + b} + |x|]$.
    This is positive and strictly decreasing to zero when $|x| \to \infty$.
    At the same time, the right-hand side strictly increases when $|x| \to \infty$: this means that there is at most one point $t^*$ such that $|x| < t^*$, where $t^*$ is found by solving Eq.~\eqref{eq:ineq3} as an equality.

	Define $q := \sqrt{(t^*)^2 + d} + t^*$ and $p := \sqrt{(t^*)^2 + b} + t^*$.
	We can verify the following equality holds:
	\begin{equation}
		t^* = \frac{1}{2} \qty(q - \frac{d}{q}) = \frac{1}{2} \qty(p - \frac{b}{p}).
	\end{equation}
	The equality point corresponds to $|b|/p = cq$, which can be used to eliminate $p$ and find
	\begin{equation}
		q^2 = \frac{|b|+cd}{c(1 + \mathrm{sgn}(b)c)}.
	\end{equation}
	The solution is then found by writing $t^* = (q^2 - d)/2q$.
	The condition $q^2 > |d|$ is the requirement that both $t^* > 0$ and $\sqrt{(t^*)^2 + d} = \frac{1}{2}(q + d/q) > 0$.
	Moreover, the same requirement for $p$ writes $p^2 > |b|$, which becomes $q^2 < |b|/c^2$ since $p = |b|/cq$.
\end{proof}

\begin{lemma} \label{lemma:ineq4}
    Let $b, c, d \in \mathbb{R}$ such that $c > 0$ and $b, d \neq 0$.
	Assuming $x^2 + b \geq 0$ and $x^2 + d \geq 0$, the inequality
	\begin{equation}
		\sqrt{x^2 + b} + |x| < c \, \mathrm{sgn}(d) \qty[\sqrt{x^2 + d} - |x|]
	\end{equation}
	admits solutions only when
	\begin{equation}
		\frac{|b|}{c} < \frac{c|d|+b}{c + \mathrm{sgn}(d)} < c|d|,
	\end{equation}
	and in this case the solution set is
	\begin{equation}
		|x| < \frac{|b - c^2 d|}{2 \sqrt{c(c + \mathrm{sgn}(d))(c|d|+b)}}.
	\end{equation}
\end{lemma}
\begin{proof}
	We define $t^*, p, q$ as in Lemma~\ref{lemma:ineq3}.
	This time, $p$ and $q$ are related by $p = c |d|/q$, and
	\begin{equation}
		q^2 = \frac{cd^2 (c + \mathrm{sgn}(d))}{c|d| + b}.
	\end{equation}
	We recover the stated result from $t^* = (q^2-d)/2q$.
	As before, the additional condition is obtained by requiring $t^* > 0$, $\sqrt{(t^*)^2 + d} > 0$ and $\sqrt{(t^*)^2 + b} > 0$.
\end{proof}


\bibliography{biblio.bib}